\documentclass[letterpaper,12pt]{article}
\pdfoutput=1 

\usepackage{jheppub} 
\usepackage{amsmath,amsfonts,amssymb}
\usepackage{psfrag}
\usepackage{enumerate}
\usepackage{mathrsfs}
\usepackage{graphicx}
\usepackage{wrapfig}
\usepackage{xcolor}
\usepackage{caption}
\usepackage{amsthm}

\numberwithin{equation}{section}

\makeatletter
\def\@fpheader{\relax}
\makeatother

\newcommand{\be}{\begin{equation}}
\newcommand{\ee}{\end{equation}}
\newcommand{\beq}{\begin{eqnarray}}
\newcommand{\eeq}{\end{eqnarray}}
\newtheorem{mydef}{Definition}

\def\[{\left [}
\def\]{\right ]}
\def\({\left (}
\def\){\right )}

\def\r2{\sqrt{2}}

\newcommand{\bbibitem}[1]{\bibitem{#1}\marginpar{#1}}

\def\Label#1{\label{#1}%
  \smash{\hbox to0pt{\raise1ex\hbox{\tiny[#1]}\hss}}}
\def\noLabels{\let\Label=\label}
\def\nobbibitem{\let\bbibitem=\bibitem}

\subheader{MIT-CTP/5106}

\title{A Modular Sewing Kit for Entanglement Wedges}

\author[a]{Bartlomiej Czech,}
\emailAdd{bartlomiej.czech@gmail.com}
\affiliation[a]{Institute for Advanced Study, Tsinghua University, Beijing 100084, China}
\author[b]{Jan de Boer,}
\affiliation[b]{Institute for Theoretical Physics and Delta Institute for Theoretical Physics, University of Amsterdam, PO Box 94485, 1090GL, Amsterdam, The Netherlands}
\emailAdd{j.deboer@uva.nl}
\author[c]{Dongsheng Ge,} 
\affiliation[c]{Laboratoire de Physique de l'\'Ecole Normale Sup\'erieure, ENS, Universit\'e PSL, CNRS, Sorbonne Universit\'e, Universit\'e Paris-Diderot, Sorbonne Paris Cit\'e, 24 rue Lhomond, 75005 Paris, France}
\emailAdd{dongsheng.ge@ens.fr}
\author[d]{Lampros Lamprou}
\affiliation[d]{Center for Theoretical Physics, Massachusetts Institute of Technology, Cambridge, MA 02139-4307, USA}
\emailAdd{llamprou@mit.edu}

\abstract{We relate the Riemann curvature of a holographic spacetime to an entanglement property of the dual CFT state: the Berry curvature of its modular Hamiltonians. The modular Berry connection encodes the relative bases of nearby CFT subregions while its bulk dual, restricted to the code subspace, relates the edge-mode frames of the corresponding entanglement wedges. At leading order in $1/N$ and for sufficiently smooth HRRT surfaces, the modular Berry connection simply sews together the orthonormal coordinate systems covering neighborhoods of HRRT surfaces. This geometric perspective on entanglement is a promising new tool for connecting the dynamics of entanglement and gravitation.}

\begin{document}
\noLabels 
\nobbibitem 

\maketitle
\flushbottom
\tableofcontents

\section{Entanglement as a Connection}

Subregion duality has taught us that the physics in a bulk entanglement wedge, i.e. its geometry, quantum state and dynamics of local quantum fields, can be recovered from the state $\rho$ and operator algebra $\mathcal{A}(\mathcal{O})$ of its dual CFT subregion \cite{Dong:2016eik, Almheiri:2014lwa, Harlow:2016vwg, Faulkner:2017vdd}. The cornerstone of this important insight was the equivalence between the modular Hamiltonians on the two sides of the duality \cite{Faulkner:2013ana, Jafferis:2014lza, Jafferis:2015del} within the code subspace \cite{Harlow:2016vwg}
\begin{equation}
H_{\text{mod}}^{\text{CFT}} = \frac{A}{4G_N} +H_{\text{mod}}^{\text{bulk}} \label{JLMS}
\end{equation}
with the modular Hamiltonians defined as $H_{\text{mod}}=-\log{\rho}$ and $A$ the area operator of the HRRT surface \cite{Ryu:2006bv, Hubeny:2007xt} bounding the entanglement wedge. 

In this paper, we utilize relation (\ref{JLMS}) to make precise how boundary entanglement `builds' the bulk spacetime \cite{VanRaamsdonk:2010pw,Maldacena:2013xja} by sewing together entanglement wedges to produce its global geometry. In ordinary differential geometry, spacetime is constructed by consistently gluing  small patches of Minkowski space, the local tangent spaces of a base manifold. Central to this task is the \emph{spacetime connection} that relates the Lorentz frames of nearby tangent spaces and endows spacetime with its curvature. Adopting this spirit, we explain how holographic spacetimes are assembled by the set of entanglement wedges by means of a geometric connection, which we propose is determined microscopically by the entanglement structure of the dual CFT state. The curvature of this entanglement connection reflects the bulk curvature in a way we make precise in Section \ref{sec:bulk}.

Our central idea is to treat entanglement as a quantum notion of connection between subsystems \cite{Czech:2018kvg}. All correlation functions within a CFT subregion $A$ with modular operator $H_{\text{mod}}^A$ are invariant under the unitary evolution generated by modular zero-modes $Q_i^A$:
\begin{equation}
\[Q_i^A, H_{\text{mod},A}\]=0
\end{equation}
where $i$ is indexing the zero-mode subalgebra. For a physicist with access only to $A$, these symmetries of her local state translate to a freedom of choice of her overall zero-mode frame. On the other hand, entanglement in the global CFT state renders the \emph{relative zero-mode frame} of different modular Hamiltonians physical, establishing a map between the algebras localized in different subregions, as we explain in Section \ref{subsec:spins}. Our proposal is to think of this zero mode ambiguity of subregions as a \emph{gauge symmetry in the space of modular Hamiltonians}. The relative modular frame is then encoded in the connection on the relevant bundle ---the modular Berry connection \cite{Czech:2017zfq} which we define for arbitrary states in Section \ref{subsec:CFTconnection}.

The bulk meaning of this zero-mode ambiguity of CFT subregions follows from relation (\ref{JLMS}). At leading order in $G_N$, bulk modular zero modes consist of large diffeomorphisms that do not vanish at the HRRT surface. These are the gravitational edge modes \cite{Donnelly:2016auv} or asymptotic symmetries \cite{Strominger:2017zoo, Hawking:2016sgy} of the extremal surface and, as we show in Section \ref{subsec:bulkzeromodes}, they consist of internal diffeomorphisms and local boost transformations on the normal 2-D plane. While the edge-mode frame for every given wedge can be chosen at will, the bulk spacetime allows us to compare frames of different wedges. When the extrinsic curvature of the HRRT surface is small compared to the Riemann curvature of the bulk spacetime, the bulk modular connection becomes the \emph{relative embedding} of the local coordinate systems about the surfaces, which is a central result in our paper (\ref{bulkconnectionoperator})$-$(\ref{bulkconnection}). The curvature of this connection includes the bulk Riemann tensor as one of its components, as we demonstrate in detail in Section \ref{subsec:bulkconnection} and \ref{subsec:bulkcurvature}.

The gravitational connection of Section \ref{sec:bulk} and the CFT connection of Section \ref{sec:CFT} encode the relations of modular Hamiltonians on the two sides of AdS/CFT and are constructed by identical sets of rules. By virtue of (\ref{JLMS}) we, therefore, propose in Section \ref{sec:conjecture} that they are related by duality. This provides a direct holographic link between the bulk curvature and the Berry curvature for the modular Hamiltonians of the CFT state. We conclude with a discussion of a number of conceptual and technical applications of the tools developed in this work.

\section{Modular Berry Connection}
\label{sec:CFT}

\subsection{A toy example}
\label{subsec:spins}
The central idea underlying this work is that entanglement plays the role of a connection for subsystems of a quantum state \cite{Czech:2018kvg}. In close analogy to the ordinary geometric connection of General Relativity which relates the Lorentz frames of nearby tangent spaces, the structure of entanglement defines the relation between the Hilbert space bases of different subsystems. 

The simplest illustration of this idea involves a system of two qubits $A$ and $B$ in a maximally entangled state:
\begin{equation}
|\psi\rangle_{AB}= \sum_{ij} W_{ij} |i\rangle_A |j\rangle_B \label{qubitstate}
\end{equation}
The reduced density matrix of each qubit is maximally mixed. Both $\rho_A$ and $\rho_B$ are invariant under unitary transformations on the respective Hilbert space, which translates to a symmetry of expectation values for operators localized in $A$ or $B$:
\begin{align}
\langle \sigma_A^i \rangle &= \langle U_A^{\dagger} \sigma_A^i U_A \rangle \nonumber\\
\langle \sigma_B^i \rangle &= \langle U_B^{\dagger} \sigma_B^i U_B \rangle
\end{align}
Here $U_A,U_B\in SU(2)$ and $\sigma_{A,B}^i$ ($i=x,y,z$) are the Pauli operators that generate the algebra of observables for the corresponding qubit. Each qubit is, therefore, endowed with a `local' $SU(2)$ symmetry; in the absence of any external system of reference, the choice of the local unitary frame for $A$ or $B$ is simply a matter of convention.

Due to the entanglement of the two qubits in $(\ref{qubitstate})$, however, expectation values of $\sigma_A^i \sigma_B^j$ are not invariant under independent unitary rotations $U_A$ and $U_B$. This reflects the fact that the global state fixes the \emph{relative} unitary frame of the subsystems. More precisely, $|\psi\rangle_{AB}$ defines an anti-linear map between the two Hilbert spaces
\begin{equation}
|i\rangle_A \rightarrow |\tilde{i}\rangle_B= \,_A\langle i |\psi\rangle_{AB} = \sum_{j}W_{ij}|j\rangle_B
\label{entanglementmap}
\end{equation}
that can also be expressed as an anti-linear map between the operators on the two Hilbert spaces
\begin{align}
\sigma_A |i\rangle_A \rightarrow \tilde{\sigma}_B |\tilde{i}\rangle_B = & \,_A\langle i| \sigma^\dagger_A |\psi\rangle_{AB} \,, \,\,\,\,\forall |i\rangle_A \nonumber\\
\Rightarrow \,\,\,\tilde{\sigma}_{B, ij} = & \,\, W_{ki}\, \sigma^*_{A, kl} \,W^{-1}_{jl} \label{operatormap}
\end{align}
The operators $\tilde{\sigma}_B$ are a simple example of the mirror operators of $\sigma_A$ as discussed in \cite{Papadodimas:2013jku}, with $|\psi\rangle_{AB}$ a cyclic and separating vector for the algebra of operators acting on subsystem $A$.

\begin{figure}
        \centering
        \includegraphics[width=0.45\textwidth]{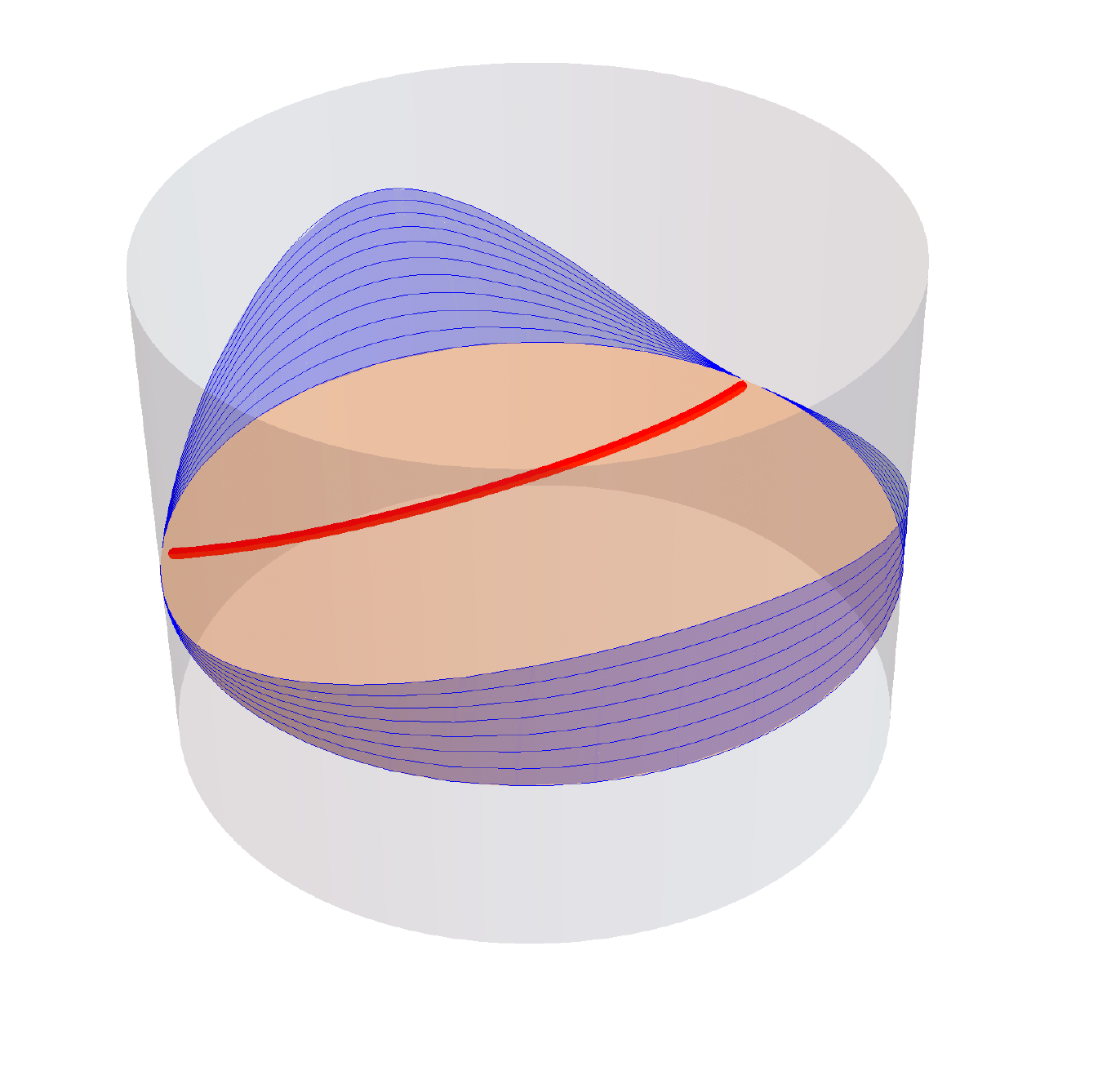}
        \qquad \quad
        \includegraphics[width=0.45\textwidth]{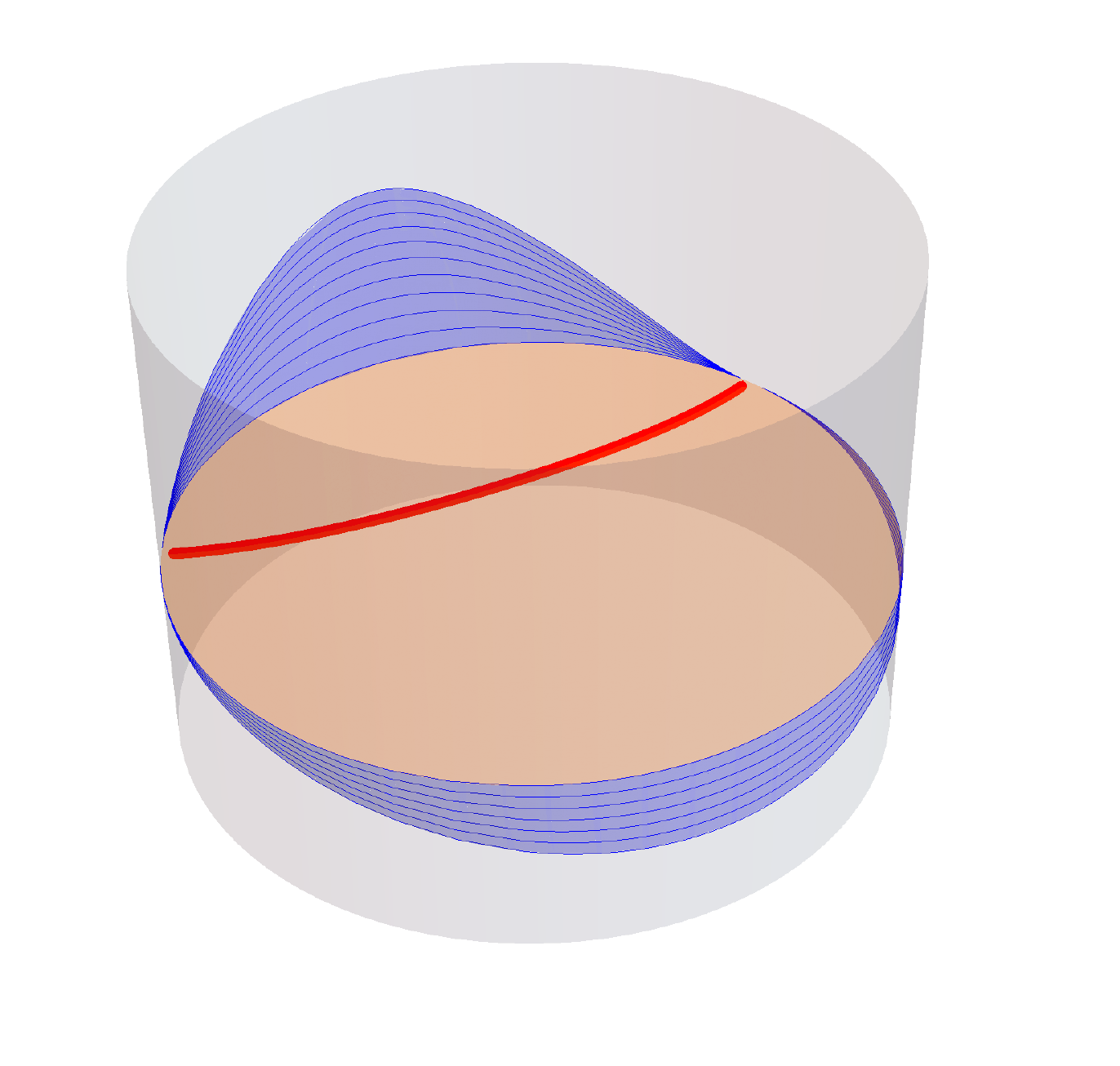}
       \caption{
       A holographic representation of eqs.~(\ref{qubitstate}-\ref{Wtransform}): the global state $W_{ij}$ of a bipartite holographic CFT is prepared by a tensor network that fills a spatial slice of the bulk spacetime (orange). The division of the CFT is illustrated with a red line that cuts through the bulk. The panels show two general examples of `gauge transformations' of `Wilson line' $W$. The focus of this paper will be on those gauge transformations, which localize on HRRT surfaces.
       }
        \label{tnconnection}
\end{figure}

It follows from definition (\ref{entanglementmap}) that the map between the two Hilbert spaces transforms under the action of a local $SU(2)$ symmetry on each qubit as:
\begin{equation}
W_{ij} \rightarrow \,  U^{\dagger}_{A,\, ik}W_{kl}\, U_{B,\, lj} \label{Wtransform}
\end{equation}
By virtue of (\ref{entanglementmap}), (\ref{operatormap}) and (\ref{Wtransform}), the matrix $W_{ij}$ can be interpreted as an open Wilson line between the Hilbert spaces of the two qubits, with a form dictated by the pattern of entanglement. From a heuristic ER=EPR viewpoint, $W_{ij}$ can be thought of as the gravitational Wilson line threading the quantum wormhole connecting the qubits \cite{Maldacena:2013xja}. 


Our main interest is in applying these observations to holographic duality; see Fig.~\ref{tnconnection}. If we divide a holographic CFT into two subregions $A$ and $B$, a global pure state can likewise be represented as a matrix $W_{ij}$ that is analogous to (\ref{qubitstate}). We can think of this matrix as being prepared by a tensor network that fills a spatial slice of the bulk spacetime. Under changes of bases in $A$ and in $B$, the matrix also transforms as in (\ref{Wtransform}), i.e. as a Wilson line. What are the corresponding Wilson loops? Are they non-trivial and what feature of the bulk spacetime do they probe? We will answer these questions in the body of the paper. One highlight is that the holonomies of the entanglement connection
probe the curvature of the dual spacetime. We interpret this as an indication that the pattern of entanglement of subsystems and the pattern of physical Wilson line dressing in gauge theories ought to be considered on equal footing.

\begin{figure}
        \centering
        \includegraphics[width=0.96\textwidth]{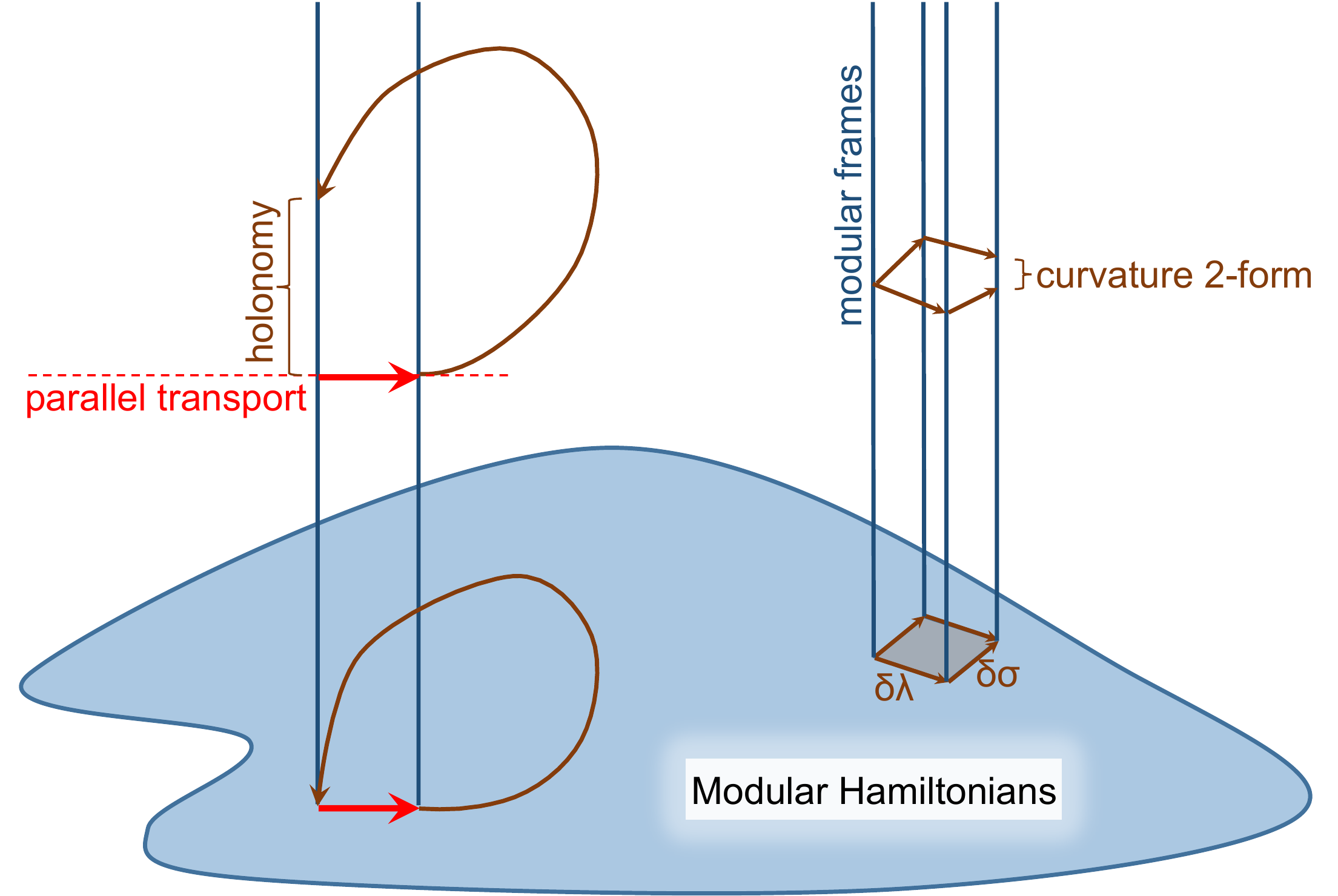}
              \caption{The fiber bundle studied in this paper. The base comprises different modular Hamiltonians and the fibers are modular zero mode frames.}
        \label{bundle}
\end{figure}

The remainder of this section is devoted to formulating the quantum notion of connection when the subsystems of interest are subregions of a CFT, in arbitrary states. As in all geometric problems that involve a connection, the correct mathematical formalism here is that of fiber bundles. A reminder of the relevant concepts from differential geometry, as well as a description of the fiber bundle at hand, is given in Fig.~\ref{bundle}. 

\subsection{Gauging the modular zero modes}
\label{subsec:CFTconnection}
Every subregion $A$ of a CFT selects an algebra of operators $\mathcal{A}_A$ that is localized in it, and a modular Hamiltonian $H_{\text{mod}}$ which encodes the reduced state in $A$ via $H_{\text{mod}}=-\log \rho_A$. 
Strictly speaking, density matrices are not well-defined objects in quantum field theory and become meaningful only in the presence of a UV cutoff. In contrast, the sum of the modular Hamiltonians of $\mathcal{A}_A$ and its commutant $\mathcal{A}_{\bar{A}}$ is well-defined in the continuum and any rigorous construction should directly refer these two-sided operators. 

We emphasize that our discussion is inherently two-sided. In order to postpone a few minor subtleties for conceptual clarity, however, we choose to phrase our initial presentation in terms of single-sided $H_{\text{mod}}$s and comment on its two-sided version in subsection \ref{subsec:2sidecomment}.

\paragraph{Modular zero modes as local symmetries.} Hermitian operators $Q^A_i$ obeying 
\begin{equation}
\[Q^A_i, H_{\text{mod},A}\]=0 \label{CFTzeromode}
\end{equation}
are called \emph{modular zero modes}. The unitary flow generated by $Q_i$  defines an automorphism of $\mathcal{A}_A$ that is a \emph{symmetry} of subregion $A$: The transformation 
\begin{equation}
O\rightarrow U^{\dagger}_Q(s_i)\, O \,U_Q(s_i) \,\,\,\,\,\, \forall\, O\in\mathcal{A}_A\,,
\end{equation}
where $U_Q= e^{-i\sum_i Q_i s_i}$, maps the algebra into itself while preserving the expectation values of all of its elements in the given state. As a result, physical data localized in a subregion carry no information about its overall zero mode frame. This local ambiguity is, of course, irrelevant for all measurements or computations restricted to $A$. It is a gauge freedom, which spans the vertical (fiber) directions of our bundle.

A useful way of describing the zero-mode ambiguity is by switching to a `Schr{\"o}dinger picture.' The modular Hamiltonian is a Hermitian operator on the CFT Hilbert space, so it can be decomposed as
\begin{equation}
H_{\text{mod}}=U^{\dagger} \Delta U \label{Hmoddecomp},
\end{equation}
where a diagonal matrix $\Delta$ encodes the spectrum and a unitary $U$ selects the basis of eigenvectors. Transformations generated by $Q_i$ preserve the form of $H_{\text{mod}}$ and, as a result, the basis $U$ in (\ref{Hmoddecomp}) is only determined up to a gauge transformation consisting of right multiplication by $U_Q$:\footnote{In other words, there is an equivalence class of CFT bases defined by $U\sim UU_Q$, in which $H_{\text{mod}}$ has identical matrix elements.}
\begin{equation}
U\rightarrow U'= UU_Q. \label{Hgaugetransformation}
\end{equation}

\begin{figure}
        \centering
        \includegraphics[width=0.80\textwidth]{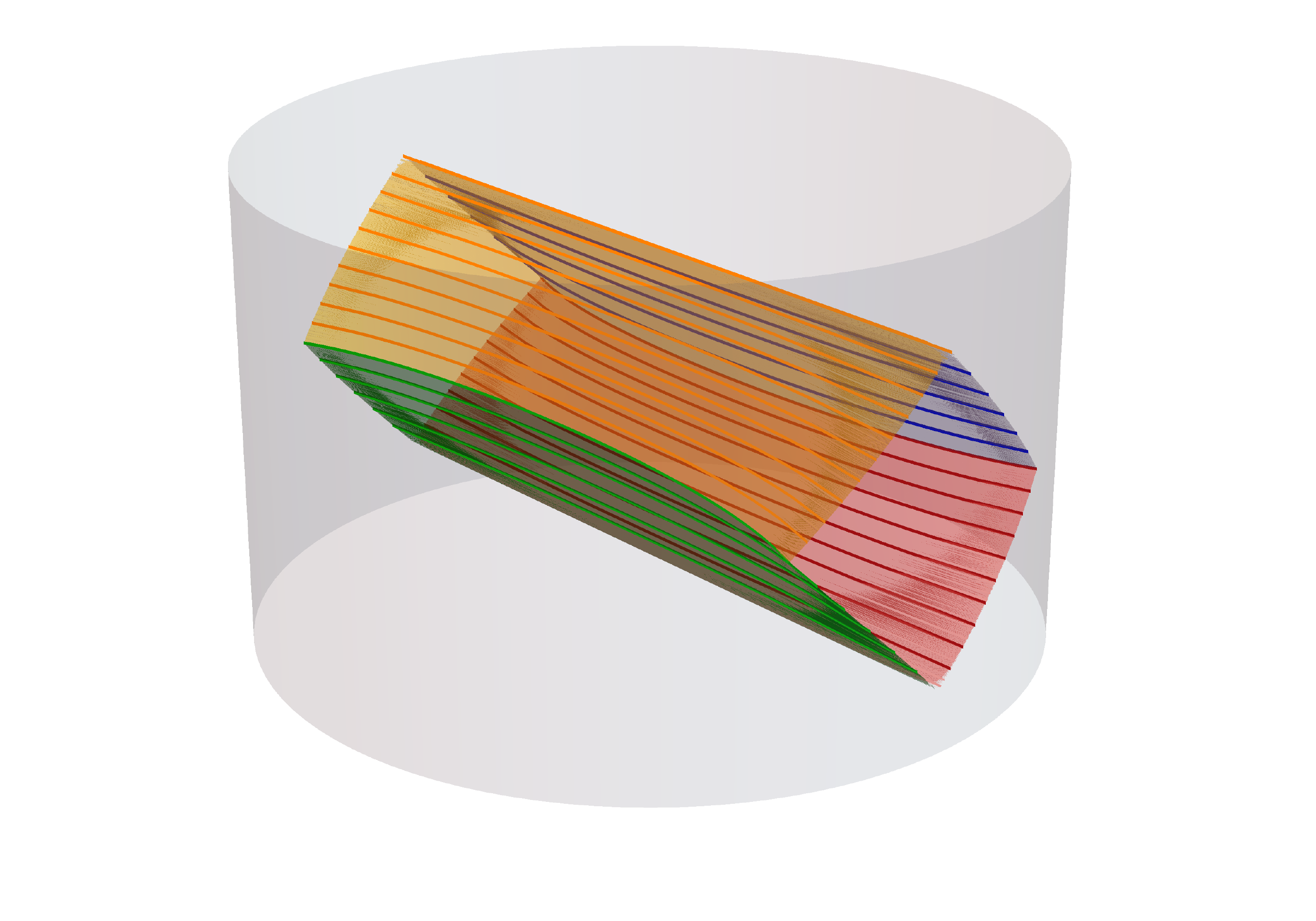}
              \caption{A closed trajectory in the space of CFT regions. To avoid clutter and to clarify the holographic application, here we display the family of corresponding RT surfaces in the bulk of AdS.}
        \label{RTfamily}
\end{figure}

\paragraph{Modular Berry connection as the relative zero-mode frame.} Consider now a continuous family of connected CFT subregions parametrized by $\lambda$ and their modular Hamiltonians $H_{\text{mod}}(\lambda)$. (See Fig.~\ref{RTfamily}.) The relations between these modular Hamiltonians can be conveniently expressed in terms of two families of operators: One describing the change of spectrum and another the precession of the basis as we vary $\lambda$. In particular, using decomposition (\ref{Hmoddecomp}), the $\lambda$-derivative of $H_{\text{mod}}(\lambda)$ is organized as
\begin{equation}
\dot{H}_{\text{mod}} = [ \dot{U}^{\dagger}U, H_{\text{mod}} ] +U^{\dagger} \dot{\Delta} U\label{Hmoddot},
\end{equation}
where $\dot{\,}\equiv \partial_{\lambda}$ and we have suppressed the $\lambda$-dependence of all operators for clarity. This shape-derivative derivative of $H_{\text{mod}}$ may cause some discomfort to the careful reader, since the density matrices of different subregions formally live in different Hilbert spaces. For an infinitesimal transformation of the region's boundary, however, this computation is in fact under control, as was shown in \cite{Faulkner:2016mzt}. The trick is to think of the shape deformation as sourcing a stress-tensor insertion at the subregion's boundary. The calculation requires a delicate treatment of the cutoff but it yields sensible results both in the CFT \cite{Faulkner:2016mzt} and holographically \cite{Lewkowycz:2018sgn}. In case this comment does not alleviate the reader's distress, we emphasize again that the discussion can be entirely expressed in terms of the full modular operators $H_{\text{mod}}(\lambda)+H_{\text{mod}}(\lambda^c)$, with $\lambda^c$ the complement of region $\lambda$, which are well-defined operators on the full CFT Hilbert space, and refer them to subsection \ref{subsec:2sidecomment} for a detailed comment.

The second term on the right hand side of (\ref{Hmoddot}) encodes the change in the spectrum of $H_{\text{mod}}$ and, since $[U^{\dagger} \dot{\Delta} U, H_{\text{mod}}]=0$, it belongs to the local algebra of modular zero modes. We can isolate this spectrum changing piece by introducing a projector $P^{\lambda}_0$ onto the zero-mode sector of $H_{\text{mod}}(\lambda)$. The latter can formally be constructed in terms of modular flow\footnote{This formula should be taken with a grain of salt since there can be Hermitian eigen-operators of the modular Hamiltonian $[H_{\text{mod}},V]=\kappa V$ which necessarily have imaginary eigenvalues, leading to exponential contributions to the integral (\ref{zeromodeprojection}).}
\begin{equation}
P^{\lambda}_0[V] \equiv \lim_{\Lambda \rightarrow \infty} \frac{1}{2\Lambda} \int_{-\Lambda}^{\Lambda} ds \,\,e^{iH_{\text{mod}}(\lambda) s}\, V e^{-iH_{\text{mod}}(\lambda) s} \label{zeromodeprojection}
\end{equation}
or in a Hilbert space representation simply as:
\begin{equation}
P^{\lambda}_0[ V] \equiv \sum_{E,q_a,q_a'} |E,q_a\rangle \langle E, q_a| V |E,q'_a\rangle \langle E,q'_a| ,
\label{zeromodeprojectionHS}
\end{equation}
where $|E,q_a\rangle$ are simultaneous eigenstates of $H_{\text{mod}}(\lambda)$ and a commuting set of zero modes $Q_a$, with eigenvalues $E$ and $q_a$ respectively. For systems with finite-dimensional Hilbert spaces, the zero-mode projector takes another useful form. Hermitian operators on a Hilbert space $\mathcal{H}$ with $\text{dim}(\mathcal{H})=D$ form a vector space and $\{\mathbb{I}, T_i\}$, where $T_i$ the $SU(D)$ generators, form a complete basis which is orthonormal with respect to the Frobenius inner product $(T_i |T_j) = \frac{1}{D} \text{Tr}\left[ T_i T_j\right] = \delta_{ij}$. One can, therefore, find an orthonormal basis of modular zero modes $\{Q_i\}$, $[Q_i, H_{\text{mod}}]=0$ and $(Q_i |Q_j)=\delta_{ij}$ and define the projector:
\begin{equation}
P_0[V]= \sum_i (Q_i|V) Q_i = \sum_i \frac{1}{D}\text{Tr}[Q_i V] Q_i
\end{equation}

An application of $P^{\lambda}_0$ on both sides of eq. (\ref{Hmoddot}) then equates the spectrum changing operator to the zero mode component of $\dot{H}_{\text{mod}}$:
\begin{equation}
U^{\dagger} \dot{\Delta} U=P_0^{\lambda} [\dot{H}_{\text{mod}}(\lambda) ] \label{deltadot}
\end{equation}
The operator $ \dot{U}^{\dagger}(\lambda)U(\lambda)$ in (\ref{Hmoddot}), in turn, encodes the change of basis accompanying an infinitesimal shape variation of the region. Combining eq. (\ref{Hmoddot}) and (\ref{deltadot}) the relative basis operator is defined as the solution to equation:
\begin{equation}
\dot{H}_{\text{mod}}- P_0^{\lambda}[\dot{H}_{\text{mod}}] = [ \dot{U}^{\dagger}U, H_{\text{mod}}(\lambda) ] \label{relativebasis}
\end{equation}
As is apparent, (\ref{relativebasis}) fixes $ \dot{U}^{\dagger}(\lambda)U(\lambda)$ only up to addition of zero modes. This ambiguous zero mode component is precisely the information we seek and it leads us to introduce the modular Berry connection:

\begin{mydef} Consider the space of CFT subregions $\mathcal{K}$, parametrized by a set of coordinates  $\lambda^i $.\footnote{The index $i$ here can be discrete or continuous. For subregions of quantum field theories in $d$ spacetime dimensions, which is our main focus in this paper, $\lambda^i$ stands for the shape and location of the subregion's boundary.}  The \textbf{modular Berry connection} is a 1-form in $\mathcal{K}$ that encodes the \textbf{relative zero mode frame} of infinitesimally separated modular Hamiltonians $H_{\text{mod}}(\lambda^i)$ and $H_{\text{mod}}(\lambda^i+\delta\lambda^i)$ and is given by:
\begin{equation}
\Gamma(\lambda^i, \delta\lambda^i)= P^{\lambda}_0[\partial_{\lambda^i}{U}^{\dagger}U ]\,\delta \lambda^i, 
\label{modularconnection}
\end{equation}
where $P^{\lambda}_0$ is the projector onto the zero-mode sector of $H_{\text{mod}}(\lambda^i)$ given by (\ref{zeromodeprojection}) or (\ref{zeromodeprojectionHS}).
Under a $\lambda$-dependent gauge transformation (\ref{Hgaugetransformation}), the connection (\ref{modularconnection}) transforms as:
\begin{equation}
U(\lambda)\rightarrow U'(\lambda)= U(\lambda)U_Q(\lambda) \,\Rightarrow \,\, \Gamma \rightarrow \Gamma'=U^{\dagger}_Q \,\Gamma \, U_Q - {U}_Q^{\dagger} \partial_{\lambda^i}U_Q\,\delta \lambda^i
\end{equation}
\end{mydef}
For the readers who may find expression (\ref{modularconnection}) for the Berry connection unfamiliar or confusing, in Appendix \ref{appendix:Berry} we include a short illustration of how (\ref{modularconnection}) reduces to the standard Berry connection \cite{Berry:1984jv,Wilczek:1984dh} when applied to a family of pure states.

\paragraph{Modular parallel transport and holonomies.} Granted a connection on a bundle, we can define a covariant derivative
\begin{equation}
D_{\lambda}^{(r)} = \partial_{\lambda} +\Gamma^{(r)},
\end{equation}
where $\Gamma^{(r)}$ is an appropriate representation of connection (\ref{modularconnection}). This covariant derivative generates \emph{parallel transport}. Any charged object, parallel transported along a closed loop $\mathcal{C}$, returns to its starting point transformed by the holonomy of $\mathcal{C}$. 

Consider now a continuous 1-parameter family of modular Hamiltonians of a QFT state, $H_{\text{mod}}(\lambda)$, $\lambda\in[0,1]$, which forms a closed loop $H_{\text{mod}}(0)=H_{\text{mod}}(1)$. The operator $U(\lambda)$ from eq.~(\ref{Hmoddecomp}), which encodes the local choice of basis in every subregion, is charged under the zero modes with transformation rule (\ref{Hgaugetransformation}). Therefore, we can compute the modular Berry holonomy of our closed loop by solving the transport problem for $U$. 

Parallel transport of $U(\lambda_0)$ assigns a basis $\tilde{U}(\lambda)$ to the modular Hamiltonians $H_{\text{mod}}(\lambda)$ for all $\lambda\in [0,1]$, with the initial conditon $\tilde{U}(\lambda_0)=U(\lambda_0)$. For an infinitesimal step $\delta \lambda$ away from $\lambda_0$, $\tilde{U}$ is equal to
\begin{equation}
\tilde{U}(\lambda_0 + \delta \lambda)\approx U(\lambda_0) + \delta\lambda\, D_{\lambda} U(\lambda_0)= U(\lambda_0+\delta \lambda)+U(\lambda_0)P_0^{\lambda_0} [\dot{U}^{\dagger}U]\delta\lambda \label{infinitesimaltransport}
\end{equation}
Multiplying both sides of (\ref{infinitesimaltransport}) with $\tilde{U}^{\dagger}(\lambda_0)$ from the left, we observe that the operator $V_{\delta\lambda}=\tilde{U}^{\dagger}\frac{\delta}{\delta\lambda}\tilde{U}$ that generates the parallel transport of the basis obeys the conditions:
\begin{align}
\dot{H}_{\text{mod}}-P_0^{\lambda}[\dot{H}_{\text{mod}}]&= [V_{\delta\lambda}(\lambda), H_{\text{mod}}]\nonumber\\
P_0^{\lambda}[V_{\delta\lambda}(\lambda)]&=0 \label{Vtransportdefinition}
\end{align}
Equations (\ref{Vtransportdefinition}) \emph{define} the modular Berry transport. In section \ref{subsec:CFTexamples}, we solve this transport problem in two tractable examples and compute the modular curvature.

\paragraph{What are the modular zero modes?} An important comment is in order. In a typical CFT state, the only symmetries of the modular Hamiltonian of a subregion are generated by $H_{\text{mod}}$ itself or the zero-modes of any globally conserved charges---or they are phase rotations of individual modular eigenstates. However, in anticipation of a connection to holography, it is important to recall that the equivalence of the bulk and boundary modular operators (\ref{JLMS}) proposed by JLMS \cite{Jafferis:2015del} holds within a code subspace
\begin{equation}
P_{\text{code}}H_{\text{mod}}P_{\text{code}}= \frac{A}{4G_N} +H_{\text{bulk}}
\end{equation}
as articulated in the error correction framework of \cite{Harlow:2016vwg}. In connecting our CFT discussion to the bulk we are, therefore, not interested in exact zero-modes but only in approximate ones, constructed by the requirement that they commute with the code subspace projection of $H_{\text{mod}}$
\begin{equation}
[Q_i,P_{\text{code}}H_{\text{mod}}P_{\text{code}}]=0. \label{codezeromodestext}
\end{equation}
We discuss the importance of this point in more detail in Section \ref{sec:conjecture}.

\subsection{Comment on two-sided modular Hamiltonians}
\label{subsec:2sidecomment}
Having concluded the presentation of our CFT formalism in the language of subregion modular Hamiltonians, we wish to illustrate that the construction can be phrased directly in terms of the full modular Hamiltonians of CFT bipartitions, $H_{\text{full}}(\lambda)\equiv H_{\text{mod}}(\lambda)+H_{\text{mod}}(\lambda^c)$, with $\lambda^c$ the complement of region $\lambda$. This is important because it is $H_{\text{full}}(\lambda)$, and not $H_{\text{mod}}(\lambda)$, that generate a well-defined unitary flow in continuum quantum field theories.

First note that the zero-modes of the single-sided modular Hamiltonians (\ref{CFTzeromode}) are obviously a subset of zero-modes of the full modular operator. However, $H_{\text{full}}$ has a much larger set of zero-modes $\tilde{Q}_i$. These generate unitary transformations on the entire Hilbert space that do not necessarily factorize to products of unitary operators on the two complementary subregions. Intuitively, they are transformations that are allowed to change the density matrices $\rho_{\lambda}$, $\rho_{\lambda^c}$ but preserve $\rho_{\lambda}\otimes \rho^{-1}_{\lambda^c}$. The fiber bundle associated to the full modular Hamiltonians has, therefore, a much larger gauge group than the one for the single-sided $H_{\text{mod}}$s.

Nevertheless, the modular Berry holonomies associated to a given global state $|\psi\rangle$ are identical for the two problems. The reason is that the Hilbert space vector $|\psi\rangle$ ``spontaneously breaks'' the symmetry group of  $H_{\text{full}}(\lambda)$ to the subgroup that preserves the state:
\begin{equation}
U_{\tilde{Q}_i}|\psi\rangle = |\psi\rangle \label{symmetrybreaking}
\end{equation}
As a result, parallel transport will only generate holonomies valued in the much smaller subgroup of zero-modes (\ref{symmetrybreaking}) which is shared between the two-sided and single-sided modular operators. The vanishing of the Berry curvature components along the extra zero-mode directions of $H_{\text{full}}$ implies that there is a globally consistent gauge in which the relevant projection of the connection vanishes everywhere and the computation reduces to the one presented in the previous section.

\subsection{Modular Berry holonomy examples}
\label{subsec:CFTexamples}

\subsubsection{CFT$_2$ Vacuum} 
\label{cftvac}
We now put our definition (\ref{modularconnection}) to work and explicitly compute the modular curvature in a tractable, illustrative example: the vacuum of a CFT$_2$ on a circle. This was computed previously in \cite{Czech:2017zfq} by exploiting the geometry of the space of CFT intervals, or kinematic space \cite{Czech:2016xec,deBoer:2016pqk}. This subsection establishes the consistency of the general rules proposed here with the results of \cite{Czech:2017zfq}. 

The (two-sided) vacuum modular Hamiltonian of an interval is an element of the conformal algebra. The global $SO(2,2)$ symmetry algebra of a CFT$_2$ decomposes to a pair of commuting $SO(2,1)$ subalgebras, which act on left-moving and right-moving null coordinates $x^+$ and $x^-$, respectively. The commutation relations are
\begin{align}
[L_0,L_1] = -L_1\,\,\,\,\,\,\,\,\,\,\,\,
&[L_0,L_{-1}] = L_{-1} \,\,\,\,\,\,\,\,\,\,\,\,
[L_1, L_{-1}] = 2 L_0 \label{conformalalgebra}
\end{align}
and similarly for $\bar{L}_i$. 

The modular Hamiltonian of the interval with endpoints at $x^\mu_L=(a^+,a^-)$ and $x^\mu_R=(b^+,b^-)$ is the generator of the boost transformation that preserves $x_L$ and $x_R$ and has the form
\begin{align}
&H_{\text{mod}}=K_+ + K_-, \nonumber
\end{align}
where $K_+$ and $K_-$ are linear combinations of $L_{-1,0,1}$ and $\bar{L}_{-1,0,1}$. Their coefficients are functions of the endpoint coordinates of the interval; we derive them in Appendix \ref{appendix:Vacuum}.

In order to compute modular Berry holonomies, we need to solve the parallel transport problem for the basis of the modular Hamiltonian. For example, given two nearby modular Hamiltonians $K_+ (a^+,b^+)$ and $K_+ (a^++da^+,b^+ )$, we need to find an operator $V_{\delta a^+}$ that solves equations~(\ref{Vtransportdefinition}). Using the explicit form of the modular Hamiltonian~(\ref{parameters}) and the conformal algebra, we find that
\begin{equation}
\partial_{a^+} K_+=\frac{1}{2\pi i} \left[  \partial_{a^+} K_+,\, K_+ \right]
\label{horizontalvacuum}
\end{equation}
so in this case parallel transport is generated by $(1/2 \pi i)\, \partial_{a^+} K_+$. More details of the calculation, as well as parallel transport along more general trajectories in the space of CFT intervals, are given in Appendix~\ref{appendix:Vacuum}.

The modular Berry curvature can now be computed straightforwardly: 
\begin{align}
R[\delta a^+, \delta b^+] &= - \frac{1}{2\pi i}\,\frac{K_+}{\sin^2\left(b^+ -a^+\right)/{2}} \nonumber\\
R[\delta a^-, \delta b^-]&= - \frac{1}{2\pi i}\,\frac{K_-}{\sin^2 {\left(b^- -a^-\right)}/{2}} \label{vacuumcurvature}
\end{align}
This exercise can also be applied to the computation of holonomies for modular Hamiltonians of ball-shaped regions in the vacuum of higher dimensional CFTs.

\subsubsection{Null deformations and modular inclusions}
The solution to the modular Berry transport becomes tractable in another interesting example: Families of modular Hamiltonians for subregions with null separated boundaries, in a CFT$_d$ vacuum. The origin of the simplification in this case is not conformal symmetry but, more interestingly, an algebraic QFT theorem for \emph{half-sided modular inclusions} \cite{Casini:2017roe}. 

Two operator subalgebras $\mathcal{A}_1$ and $\mathcal{A}_2 \subset  \mathcal{A}_1$ are said to form a \emph{modular inclusion} if modular evolution by $H_{\text{mod},1}$ maps  $\mathcal{A}_2$ into itself for all \emph{positive} modular times:
\begin{equation}
U^{\dagger}_{\text{mod},1}(s) \mathcal{A}_2 U_{\text{mod},1}(s) \subset \mathcal{A}_2 \,\,\,\,\, \forall s>0
\end{equation}
The half-sided modular inclusion theorem then states that the modular Hamiltonians of included algebras satisfy the commutator:
\begin{equation}
[H_2, H_1] = 2\pi i (H_2 -H_1) \label{modularinclusion}
\end{equation}

In the CFT$_d$ vacuum, when two subregions are related by an infinitesimal null deformation $x^{\mu}_{e.s.}\rightarrow x^{\mu}_{e.s.}+u^\mu(x_{e.s})$ their algebras are indeed included and (\ref{modularinclusion}) directly implies:
\begin{equation}
\[\frac{\delta H_{\text{mod}}}{\delta u(x)}, H\] = 2\pi i \,\frac{\delta H_{\text{mod}}}{\delta u(x)}
\end{equation}
The null functional derivative of the modular Hamiltonian $\frac{\delta H_{\text{mod}}}{\delta u(x)} $ is, therefore, an eigen-operator of $H$ with non-zero eigenvalue $2\pi i$ that satisfies (\ref{Vtransportdefinition}). The parallel transport operator for null deformations is then:
\begin{equation}
V_{\delta u} = \frac{1}{2\pi i} \frac{\delta H_{\text{mod}}}{\delta u(x)}
\end{equation}

\section{Entanglement Wedge Connection}
\label{sec:bulk}
\subsection{Modular zero modes in the bulk}
\label{subsec:bulkzeromodes}

The link between our CFT discussion and the bulk gravity theory is the JLMS relation \cite{Faulkner:2013ana, Jafferis:2014lza, Jafferis:2015del}. The modular Hamiltonian of a boundary subregion is holographically mapped to:
\begin{equation}
H_{\text{mod}}= \frac{A}{4G_N} +H_{\text{mod}}^{\text{bulk}} \label{JLMS2}
\end{equation}
where $A$ is the HRRT surface area operator and $H_{\text{mod}}^{\text{bulk}} $ the modular Hamiltonian of the bulk QFT state in the associated entanglement wedge. This operator equivalence holds within the subspace of the CFT Hilbert space that corresponds to effective field theory excitations about a given spacetime background, called the code subspace \cite{Harlow:2016vwg}.

An important consequence of (\ref{JLMS2}) is that, for all holographic states of interest, $H_{\text{mod}}$ admits a geometric description in a small neighborhood of the HRRT surface \cite{Faulkner:2017vdd,Faulkner:2018faa}. The area operator in Einstein gravity is identified with the Noether charge for diffeomorphisms $\zeta_{\text{mod}}^M$ that asymptote to a homogenous boost near the RT surface. Moreover, for finite energy bulk states, $H_{\text{mod}}^{\text{bulk}}$ reduces to its vacuum expression in the same neighborhood, implementing the above boost transformation on the matter fields. This renders the, generally non-local, $H_{\text{mod}}$ a geometric boost generator at the edge of the entanglement wedge.

To make our discussion concrete, we partially fix the gauge to be orthonormal to the HRRT surface
\begin{align}
g&= \Big(\eta_{\alpha\beta} +w_{\alpha\beta|\gamma}(y)\,x^{\gamma} +\mathcal{O}(x^2) \Big)dx^{\alpha}\otimes dx^{\beta} + \Big(2\sigma_{i \alpha|\beta}(y)\, x^\beta +\mathcal{O}(x^2) \Big) dx^{\alpha}\otimes dy^i \, + \nonumber\\
&+ \Big(\gamma_{ij}(y) + k_{ij|\alpha}(y)\, x^{\alpha} +\mathcal{O}(x^2) \Big) dy^i \otimes dy^j  \label{gauge}
\end{align}
where $y^{i}, \, i=2,\dots d$ is some choice of coordinates along the minimal surface directions and $x^{\alpha}, \, \alpha=0,1$ parametrize distances along two orthogonal transverse directions, with the extremal surface at $x^{\alpha}=0$. The boost generated by the CFT modular Hamiltonian in the gauge (\ref{gauge}) reads:
\begin{align}
\zeta_{\text{mod}}^\alpha &\overset{x^{\alpha}\sim 0}{\rightarrow} 2\pi \epsilon^{\alpha \beta} x_\beta   \nonumber\\
\zeta_{\text{mod}}^i &\overset{x^{\alpha}\sim 0}{\rightarrow} 0\label{modulardiffeo}
\end{align}
This approximation for the modular flow is valid within a neighborhood with size set by the normal extrinsic curvature of the HRRT surface $ x^{+}K_{ij|+}, \,x^{-}K_{ij|-}\ll 1$ where $x^{\pm}$ normal lightlike coordinates \cite{Faulkner:2017vdd}. Beyond this regime, modular flow gets modified by generically non-local contributions. Our entire discussion in this section assumes the validity of approximation (\ref{modulardiffeo}). We will discuss how the corrections restrict the regime of validity of our results in Section \ref{subsec:bulkcurvature}.

\paragraph{Bulk modular zero modes.} As is the case for the modular Hamiltonian itself, the zero modes will generally be non-local operators in the bulk wedge, defined by $\[Q_i, H_{\text{mod}}^{\text{bulk}}\]=0$. Near the extremal surface, however, due to the geometric action of $H_{\text{mod}}$ a class of zero modes will reduce to generators of spacetime transformations:
\begin{equation}
Q=\zeta^M(y^i) \partial_M + \mathcal{O}(x^{+}K_{ij|+}, \,x^{-}K_{ij|-})  \label{zeromodeop}
\end{equation}
These need to preserve the location and area of the HRRT surface and commute with the modular boost (\ref{modulardiffeo}), which translates to the condition
\begin{align}
\[\zeta, \zeta_{\text{mod}}\]^M&=\zeta^N \partial_N \zeta_{\text{mod}}^M - \zeta_{\text{mod}}^N\partial_N \zeta^M=0. \label{bulkzeromodes}
\end{align}
Moreover, we demand that the diffeomorphisms generated by (\ref{zeromodeop}) are non-trivial. In a spacetime with no boundary, all spacetime transformations have vanishing generators, as a result of the constraint equations of gravity. When boundaries exist, however, diffeomorphisms that act non-trivially on them are endowed with non-vanishing Noether charges.

An entanglement wedge has two boundaries: The standard asymptotic boundary used to define CFT correlators and the boundary selected by the HRRT surface. Large diffeomorphisms that do not vanish asymptotically give rise to the boundary conformal group and they are not relevant for us here. On the other hand, diffeomorphisms that act non-trivially on the HRRT surface have Noether charge \cite{Camps:2018wjf}
\begin{equation}
Q^{\text{Noether}}_{\zeta}=-\frac{1}{4\pi G_N}\int_{RT} \sqrt{\gamma}\, \epsilon^{\alpha\beta}\nabla_{\alpha}\zeta_{\beta}
\end{equation}
and constitute bulk zero-modes when $Q^{\text{Noether}}_{\zeta}\neq 0$ and (\ref{bulkzeromodes}) is satisfied. The modular boost is, of course, one of them, with a Noether charge equal to the area of the extremal surface in Planck units.

The symmetry group selected by the above requirements consists of diffeomorphisms along the minimal surface directions and location-dependent boosts in its normal plane. In gauge (\ref{gauge}) the zero-modes read:
\begin{align}
\zeta^\alpha\, \overset{x^{\alpha}\sim 0}{\longrightarrow}\,\,\,& \omega(y)\, \epsilon^{\alpha \beta} x_\beta \nonumber\\
\zeta^i\, \overset{x^{\alpha}\sim 0}{\longrightarrow}\,\,\,& \zeta_0^i(y) + 0\cdot x \label{bulkzeromodesexplicit}
\end{align}
where in the second line we chose to explicitly emphasize the vanishing transverse derivative of the $i-$th component, $\partial_{\alpha}\zeta^i \big|_{x^{\alpha}=0}=0$, as demanded by (\ref{bulkzeromodes}). It follows that the class of zero modes $\zeta^i\partial_i$ map the HRRT surface to itself while preserving its normal frame, a fact that will play a crucial role in section \ref{subsec:bulkconnection}.

Transformations (\ref{bulkzeromodesexplicit}) are the gravitational edge modes discussed in \cite{Donnelly:2016auv} and the analogue of the horizon symmetries of \cite{Hawking:2016sgy} where our RT surface replaces their black hole horizon. As in our CFT discussion, the vector fields (\ref{bulkzeromodesexplicit}) generate symmetries of the physics near $x^{\alpha}=0$ in a given wedge and will be treated as local gauge transformations on the space of entanglement wedges. We should note that, in general, there exist other zero-modes as well, e.g. edge-modes of bulk gauge fields, that generate extra components of the modular Berry connection. The gravitational edge-modes discussed here, however, are universally present in holographic theories and for this reason we choose to focus our discussion on them.

\subsection{Relative edge-mode frame as a connection}
\label{subsec:bulkconnection}

Consider now two entanglement wedges, $\lambda$ and $\lambda+\delta \lambda$, whose HRRT surfaces are infinitesimally separated from each other. Each wedge is equipped with a vector field $\zeta^M_{\text{mod}}(x ;\lambda)$ generating the corresponding modular flow near its HRRT surface. Moreover, each wedge comes with its own arbitrary choice of zero-mode frame, which given (\ref{bulkzeromodesexplicit}) is simply an internal coordinate system on the extremal surface and a hyperbolic angle coordinate on its transverse 2D plane. Fig.~\ref{fig:holonomy} below zooms in on a small fragment of four extremal surfaces and displays their zero-mode frames. 

The key idea now is that the geometry of the global spacetime enables us to compare the two zero mode frames. What makes this possible is the existence of diffeomorphisms  $x^M\rightarrow x^M+\xi^{M}(x)$, which map one extremal surface to the other, allowing us to relate the coordinate systems in their neighborhoods. Bulk diffeomorphisms, therefore, play the role of the relative basis operator $\dot{U}^{\dagger}U$ (\ref{relativebasis}) in our CFT discussion.

\paragraph{Mapping the modular boost generators.} It is instructive to proceed in parallel with our CFT construction of Section \ref{subsec:CFTconnection}. The $\lambda$-variation of the modular Hamiltonian in the bulk becomes the difference of the vector fields $\zeta^M_{\text{mod}}(x ;\lambda)$ and $\zeta^M_{\text{mod}}(x ;\lambda+\delta\lambda)$. As in CFT, this can generally be organized into two contributions as follows
\begin{equation}
\delta_{\lambda}\zeta_{\text{mod}}^M(x;\lambda)= \[ \xi(x;\lambda,\delta\lambda) , \zeta_{\text{mod}}(x;\lambda) \]^M +P_0^{\lambda}[\delta_{\lambda}\zeta_{\text{mod}}^M(x;\lambda)] \label{bulkmodHtransport}
\end{equation}
where $\delta_{\lambda}\zeta_{\text{mod}}^M$ is the difference between the two modular generators and $P_0$ is the bulk projector onto zero modes discussed in more detail below (see eq. (\ref{bulkconnection})). The vector field $\xi(x;\lambda,\delta\lambda)$ is a diffeomorphism rotating the basis of the modular Hamiltonian, which 
in the geometric regime is simply the local coordinate system. On the other hand, the zero-mode projection describes the change in the spectrum. Condition (\ref{bulkmodHtransport}) is the direct bulk analogue of CFT equation (\ref{relativebasis}).

Equation (\ref{bulkmodHtransport}) determines the diffeomorphism $\xi$ up to additive contributions by zero-mode transformations (\ref{bulkzeromodesexplicit}). A formal but explicit solution to the general problem can be obtained as follows. First, we introduce the transverse location $\delta x^{\alpha}=\delta x^{\alpha}(y^i;\lambda, \delta \lambda)$ of the HRRT surface $\lambda+\delta\lambda$ relative to $\lambda$. Crucially, $\delta x^{\alpha}$ is determined simply by the deformation $\delta x^{\mu}_{\partial B} $ of the boundary subregion the surface is anchored at. This follows from equation
\begin{equation}
\delta_{\delta x} K_{\alpha}= -\eta_{\alpha\beta}\gamma^{ij}\nabla_i\nabla_j \delta x^{\beta}+\gamma^{ij}R_{i(\alpha\beta)j}\delta x^{\beta}-K_{\alpha;ij}K_\beta\,^{ij}\delta x^\beta=0 \label{extremality}
\end{equation}
where $K_{\alpha;ij}=\mathcal{L}_{\alpha} g_{ij}$ is the normal extrinsic curvature and $K_\alpha = \gamma^{ij}K_{\alpha; ij}$, which ensures the new surface at $\delta x^{\alpha}$ is also extremal, as discussed in detail in \cite{Lewkowycz:2018sgn}. Eq. (\ref{extremality}) fixes the form of $\delta x^{\alpha}$ in terms of its boundary condition $\delta x^{\mu}_{\partial B}$.

Second, we recall that the vector field $\zeta_{\text{mod}}$ generates boosts on the normal 2-D plane of the HRRT surface. $\zeta_{\text{mod}}(\lambda+\delta\lambda)$ therefore needs to also have the  form (\ref{modulardiffeo}) in normal coordinates about $\lambda+\delta\lambda$. To express this requirement, we introduce a pair of normal vectors on the new surface $n_a\,^M(y;\lambda+\delta\lambda) = \delta^M_a +\delta n_a\,^M(y) +\mathcal{O}(\delta n^2)$, where $a=0,1$ with $n_a\cdot n_b=\eta_{ab}.$\footnote{There is of course no unique choice. There is a continuous family of normal vectors related by local Lorentz transformations, which will be important later on.} Imposing both conditions on $\xi(x;\lambda,\delta\lambda)$ then leads to the following solution of eq.~(\ref{bulkmodHtransport}):
\begin{align}
\xi^M (x^M;\lambda, \delta\lambda)= & -\delta^M_a\,\delta x^a -\left(\delta n_a\,^M +\Gamma^M_{ab}\,\delta x^b \right)x^a   \nonumber\\
& +\omega(y) \delta^M_a\,\epsilon^a\,_b x^b  + \delta^M_i\,\zeta_0^i(y),
\label{xigeneral}
\end{align}
where $\Gamma^M_{NK}$ are the Christoffel symbols in gauge (\ref{gauge}). A detailed derivation of $\xi(x;\lambda,\delta\lambda)$ is given given in Appendix~\ref{appendix:solution}. The expression for the diffeomorphism $\xi$ in an arbitrary gauge can, of course, be obtained simply by a change of coordinates in (\ref{xigeneral}).

The quantities $\omega(y), \zeta_0^i(y)$ are arbitrary functions of the minimal surface coordinates representing the edge-mode ambiguity in $\xi$. The arbitrariness in $\omega(y)$ is precisely our freedom in selecting a pair of orthonormal vectors $n_{\alpha}\,^M$ among the family of Lorentz equivalent pairs, as can be seen by the transformation of $\delta n_a\,^M$ under a local Lorentz boost on the surface's transverse plane:
\begin{equation}
\delta n_{\beta}\,^{\alpha} \rightarrow \delta n_{\beta}\,^{\alpha}+ \omega(y) \epsilon_{\beta}\,^{\alpha}. \label{normalboost}
\end{equation}
The zero-mode $\omega$ can, therefore, be absorbed into the definition of $\delta n_a\,^M$. The undetermined function $\zeta_0^i(y)$, in turn, expresses our right to pick the coordinate system on the surface $\lambda+\delta\lambda$ at will.

\paragraph{Bulk modular connection.} The ambiguous edge-mode part in the solution of (\ref{bulkmodHtransport}) encodes the \emph{relative zero-mode frame} of the two entanglement wedges. In order to define the bulk modular connection we, therefore, need to perform a zero-mode projection of the diffeomorphism $\xi^M$, mapping between the two coordinate systems at $\lambda$ and $\lambda+\delta \lambda$. In covariant form, the zero-mode component of the vector field $\xi$ that \emph{defines} modular connection in the bulk reads:\footnote{It is straightforward to confirm that the definition of the zero-mode projector $P[\xi]=-\Omega [\xi]\epsilon^{\alpha}\,_\beta x^\beta \partial_\alpha -Z^i [\xi] \partial_i$ satisfies $P\circ P=P$, is itself a zero-mode and it annihilates the vector Lie bracket $[\xi, \zeta_{\text{mod}}]$, as any consistent projector should.} 
\begin{align}
\Gamma(\lambda,\delta\lambda)&= \Omega [\xi ] \mathcal{L}_{\Omega} +Z^i [\xi ] \mathcal{L}_{Z^i} \label{bulkconnectionoperator}
\end{align}
where $\mathcal{L}_{\Omega}$, $\mathcal{L}_{Z^i}$ are the Lie derivatives generating the corresponding asymptotic symmetries of the HRRT surface (\ref{bulkzeromodesexplicit}), and
\begin{align}
\Omega\big[\xi(\lambda,\delta\lambda)\big] &= \frac{1}{2} \epsilon^{\alpha\beta} n_{\alpha}\,^M\,\partial_M\left( n_{\beta N}\xi^N \right)\big|_{RT} \nonumber\\
&=\frac{1}{2}\epsilon^{\alpha\beta}  \left(n_{\alpha M}\,\Delta_{\delta\lambda} n_\beta\,^M+n_{\alpha M}\,\Gamma^M_{\gamma K}\delta x^{\gamma}\, n_{\beta}\,^K\right)  \label{spinconnection}\\
Z^i\big[\xi(\lambda,\delta\lambda)\big] &= -t^i_N\xi^N\big|_{RT} \label{bulkconnection}
\end{align}
Here $n_{\alpha}\,^M(\lambda)$ ($\alpha=1,2$) are two unit normal vectors on the extremal surface $\lambda$ with \mbox{$n_{\alpha}\cdot n_\beta=\eta_{\alpha\beta}$} and $t^{iN}$ ($i=1, \ldots, d-2$) are the corresponding tangents. We also introduced $\Delta_{\delta \lambda}$ for the `internal' covariant derivative  associated to the $d-2$-dimensional diffeomorphism subgroup
\begin{equation}
\Delta_{\delta\lambda}= \delta \lambda\frac{\partial}{\partial\lambda} +Z^i[\xi] \mathcal{L}_{Z^i}\label{Deltaderivative}
\end{equation}
It is very important here that the zero-mode $\mathcal{L}_{Z^i}$ preserves the normal frame, as explained around eq. (\ref{bulkzeromodesexplicit}), and, therefore, provides a canonical map between normal vectors at different locations on the same HRRT surface, allowing the construction of the internal covariant derivative (\ref{Deltaderivative}).

Expression (\ref{spinconnection}) for the boost component of $\Gamma(\lambda,\delta\lambda)$ is,  by definition, the spin connection for the normal frame of the HRRT surface. The role of the covariant derivative $\Delta_{\delta\lambda}$ is to \emph{align the internal coordinates} of the nearby minimal surfaces before comparing the normal frames at the `same location'. Thus, the curvature of our modular Berry connection computes the bulk Riemann curvature \emph{when} the approximation (\ref{modulardiffeo}) of the modular flow is justified. We explain this proviso in more detail in the next subsection.

\subsection{Bulk modular curvature and parallel transport} 
\label{subsec:bulkcurvature}
\begin{figure}
        \centering
        \includegraphics[width=.96\textwidth]{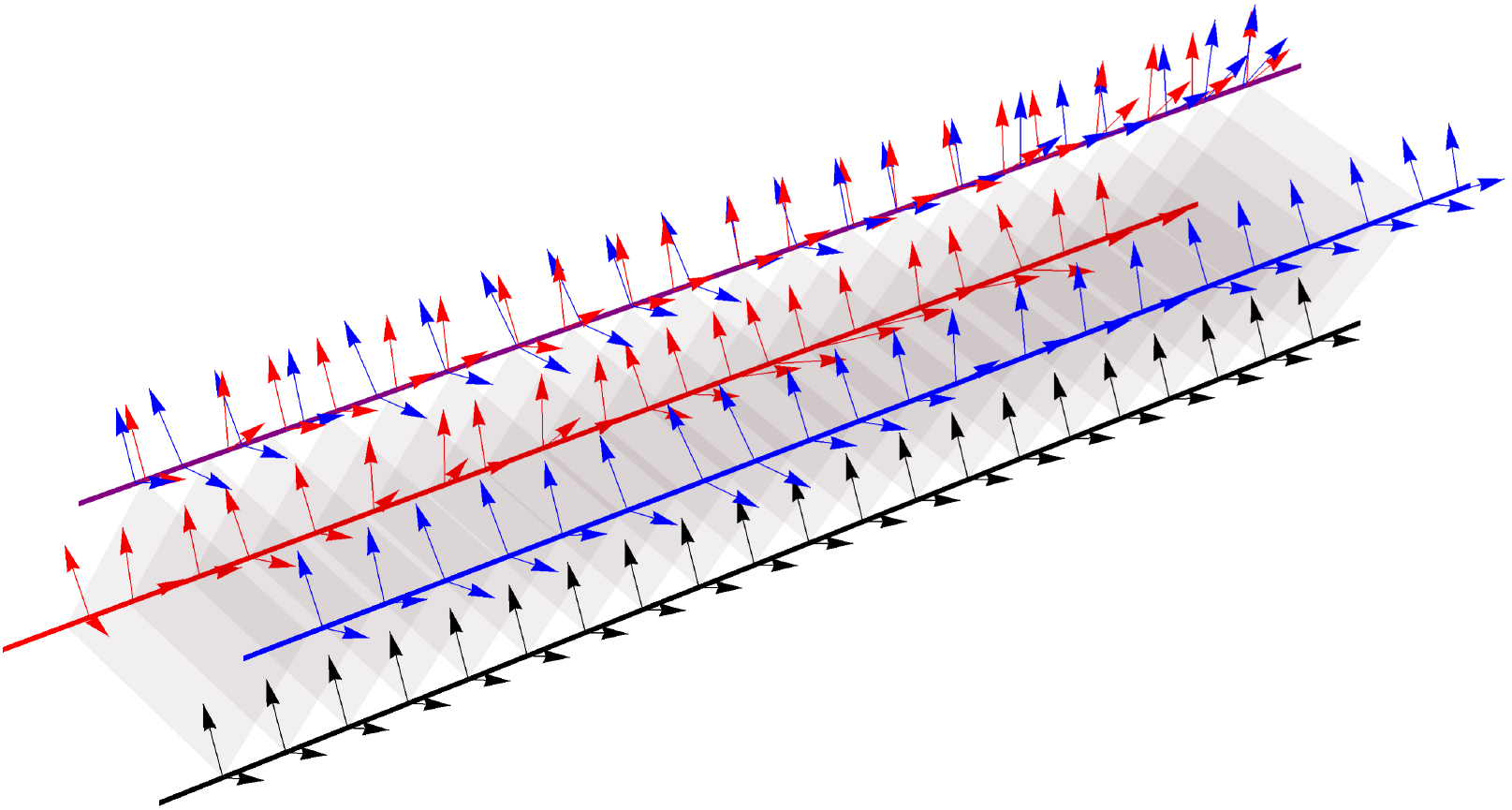}
       \caption{Modular Berry curvature in the bulk. The modular zero mode frames are marked with pairs of arrows that stand for the normal vectors $n_\alpha^M(\lambda, y)$ (which transform under orthogonal boosts); the distances between neighboring pairs reflect the extremal surface diffeomorphism frame. We parallel transport a zero mode frame from the bottom surface to the top surface along two different paths (red and blue); the mismatch between the resulting frames is the modular curvature. The mismatch between the locations of the red and blue arrows on the top is the surface diffeomorphism component of the curvature (\ref{diffcurvature}) while the mismatch between their directions is the boost component of the curvature (\ref{boostcurvature}).}
        \label{fig:holonomy}
\end{figure}

Equipped with connection (\ref{bulkconnectionoperator}), the bulk modular curvature follows from the standard definition. It reads:
\begin{align}
R_{\delta\lambda_1 \delta \lambda_2}&= \delta_{\delta \lambda_1}\Gamma(\delta\lambda_2) - \delta_{\delta \lambda_2}\Gamma(\delta\lambda_1) +[\Gamma(\delta\lambda_1),\Gamma(\delta\lambda_2)] \nonumber\\
&=\Big( \Delta_{\delta\lambda_1} \Omega(\delta \lambda_2) -\Delta_{\delta\lambda_2} \Omega(\delta \lambda_1) \Big) \mathcal{L}_{\Omega} \nonumber\\
&+ \Big( \Delta_{\delta\lambda_1} Z^i(\delta\lambda_2) -\Delta_{\delta\lambda_2}Z^i (\delta\lambda_1)   \Big)\mathcal{L}_{Z^i}\,, \label{bulkmodcurvature}
\end{align} 
where $\Delta_{\delta \lambda}$ is given by expression (\ref{Deltaderivative}). We illustrate the modular curvature in Fig.~\ref{fig:holonomy}.

The curvature (\ref{bulkmodcurvature}) can be decomposed into two contributions: the curvature of the non-abelian group of surface diffeomorphisms
\begin{equation}
R^{(Z)}_{\delta\lambda_1 \delta \lambda_2}=\Delta_{\delta\lambda_1} Z^i(\delta\lambda_2) -\Delta_{\delta\lambda_2}Z^i (\delta\lambda_1) 
\label{diffcurvature}
\end{equation} 
and the curvature of the abelian subgroup of local transverse boosts generated by $\mathcal{L}_{\Omega}$:
\begin{equation}
R^{(\Omega)}_{\delta\lambda_1 \delta \lambda_2}=\Delta_{\delta\lambda_1} \Omega(\delta \lambda_2) -\Delta_{\delta\lambda_2} \Omega(\delta \lambda_1) \label{boostcurvature}
\end{equation}
The appearance of the internal covariant derivative $\Delta_{\delta\lambda}$ in (\ref{boostcurvature}) is required by covariance because the orthogonal boosts are non-trivially fibered over the surface diffeomorphisms. 

\paragraph{Relation to the bulk curvature.} Expression (\ref{spinconnection}) for the modular Berry connection is the spin connection for the normal frame of the extremal surfaces. This immediately implies that the $\mathcal{L}_{\Omega}$ component of the curvature (\ref{boostcurvature}) directly probes the bulk Riemann curvature.

At this point it is important, however, to recall that in approximation (\ref{modulardiffeo}) of the modular flow as a boost generator we neglected terms  of order $\mathcal{O}( x^{+}K_{ij|+}, \,x^{-}K_{ij|-})$ where $K_{ij|\alpha}$ is the normal extrinsic curvature of the HRRT surface. These, generally non-local, contributions to the bulk modular Hamiltonian can generate corrections of order $\mathcal{O}(K_{ij|\alpha} \delta x^{\alpha})$ to (\ref{spinconnection}), which affect its curvature at orders $\mathcal{O}(K^2, \partial K)$. Since the modular curvature computed by na{\"\i}vely employing approximation (\ref{modulardiffeo}) at every step is the bulk Riemann curvature, our computation of (\ref{xigeneral}) is under control only when there is a hierarchy between the bulk Riemann curvature and the normal extrinsic curvature of the HRRT surface
\begin{equation}
R\gg K^2, \,\partial K. \label{regimeofvalidity}
\end{equation}
The curvature of the geometric connection (\ref{spinconnection}, \ref{bulkconnection}) is an approximation to the modular Berry connection at leading order in a $K^2/R, \partial K/R$ expansion. 

Condition (\ref{regimeofvalidity}) can be intuitively understood as follows: The geometric approximation to the bulk modular flow confines us within a distance of order $\mathcal{O}(1/K)$ from the extremal surface. If this distance is also small compared to the Riemann curvature, spacetime looks effectively flat, and the boost of the normal frame resulting from parallel transporting the surface is comparable to the corrections neglected in the approximation (\ref{modulardiffeo}). The modular Berry curvature, therefore, reliably measures the bulk curvature in the neighborhood of an RT surface when the surfaces considered obey (\ref{regimeofvalidity}).

\paragraph{Modular parallel transport.} Assuming (\ref{regimeofvalidity}), we can illustrate the modular parallel transport geometrically. Consider a family of minimal surfaces $\gamma(\lambda)$ with $\lambda \in [0,1]$ that form a closed loop $\gamma(0)=\gamma(1)$. In a neighborhood of every minimal surface $\gamma({\lambda})$ we can define a coordinate system $x_{\lambda}^M=(x_{\lambda}^{\alpha},y_{\lambda}^i)$, where $x_{\lambda}^{\alpha}$ ($\alpha=0,1$) measures distances from $\gamma(\lambda)$ along two orthogonal directions. These are simply local choices for the edge-mode frames of the corresponding entanglement wedges. As we explained in the previous section, these different localized coordinate patches are related to each other by the diffeomorphisms (\ref{xigeneral}):
\begin{equation}
x^M_{\lambda+\delta\lambda}=x^M_{\lambda}+\xi^M(x; \lambda, d\lambda).
\end{equation}
The `gluing' diffeomorphism $\xi$ is of course subject to the zero mode ambiguity, which is the focus of this paper. 

Given the connection (\ref{bulkconnectionoperator}), we can define a covariant derivative 
\begin{equation}
\nabla_{\lambda}= \frac{\delta}{\delta\lambda} +\Gamma(\lambda,\delta\lambda)
\end{equation}
which generates parallel transport. Applied to the coordinate frames $x_{\lambda}^M$, parallel transport assigns a canonical frame $\tilde{x}^M_{\lambda}$ to every surface $\gamma(\lambda)$, given an initial condition $\tilde{x}^M_{\lambda_0}=x^M_{\lambda_0}$. For an infinitesimal step $\delta\lambda$, the parallel transported frame becomes:
\begin{align}
\tilde{x}^M_{\lambda+\delta\lambda}&=\tilde{x}^M_{\lambda} +\delta\lambda\, \nabla_{\lambda}\tilde{x}^M_{\lambda} \nonumber\\
&=x^M_{\lambda+\delta\lambda} +\Big(\Omega(\lambda,\delta\lambda) \epsilon^{\alpha \beta}\tilde{x}_{\lambda \beta} \frac{\partial}{\partial \tilde{x}_{\lambda}^{\alpha}} +Z^i(\lambda,\delta\lambda) \frac{\partial}{\partial \tilde{y}_{\lambda}^i}   \Big) \tilde{x}^M_{\lambda} \nonumber\\
&=\tilde{x}^M_{\lambda} +\xi^M(\tilde{x}; \lambda,\delta\lambda) +   \left(\frac{1}{2}\epsilon^{\gamma}\,_{\delta} \partial_{\gamma}\xi^{\delta}\right)\big|_{\tilde{x}^{\alpha}=0} \delta^M_{\alpha}\epsilon^{\alpha\beta} \tilde{x}_{\lambda \beta} -\delta^M_i \xi^i\big|_{\tilde{x}^{\alpha}=0} \label{bulkparalleltransport1}
\end{align}
In the second step we used the explicit form of the zero-mode generators in the local orthonormal gauge (\ref{bulkzeromodesexplicit}) and in the third step we used the formulas (\ref{spinconnection}, \ref{bulkconnection}) for the components of the connection $\Omega, Z^i$. An application of the projector (\ref{bulkconnection}) to (\ref{bulkparalleltransport1}) reveals that the diffeomorphism $\tilde{\xi}^M(\lambda,\delta\lambda)$ generating parallel transport of the edge-mode frame  
\begin{equation}
\tilde{x}_{\lambda+\delta\lambda}^M =\tilde{x}_{\lambda}^M +\tilde{\xi}^M(\tilde{x}; \lambda,\delta\lambda)
\end{equation}
indeed has vanishing zero mode components. 

The modular parallel transport in the bulk can be summarized as a geometric flow, which at every step: 
\begin{enumerate}
\item maps between the two modular boost generators (up to zero modes),
\item is always orthogonal to the extremal surface, 
\item preserves the hyperbolic angles on the normal 2-D plane. 
\end{enumerate}

We can covariantly express these conditions as follows:
\begin{align}
\delta_{\lambda}\zeta_{\text{mod}}^M(x;\lambda)- P_0^{\lambda}[\delta_{\lambda}\zeta_{\text{mod}}^M(x;\lambda)]&= \[ \tilde{\xi}(x;\lambda,\delta\lambda) , \zeta_{\text{mod}}(x;\lambda) \]^M \\
\frac{1}{2} \epsilon^{\alpha\beta} n_{\alpha}\cdot\partial \left( n_{\beta}\cdot \tilde{\xi} \right)\big|_{RT} &=0\\
t^i\cdot \tilde{\xi}\Big|_{RT}&=0
\end{align}
They are direct bulk analogues of the CFT conditions (\ref{Vtransportdefinition}). 

Following these rules we can transport the surface around a closed loop in the space of extremal surfaces, returning to its original location in the end. A comparison of the original and transported coordinate frames in its vicinity will reveal a location-dependent boost transformation on its normal plane and a diffeomorphism of the internal coordinates. This is the bulk modular Berry holonomy. 
We saw an example of it in Fig.~\ref{fig:holonomy}, which shows the computation of the modular curvature---that is, the holonomy of an infinitesimal loop. But the picture is the same for larger loops, for example the loop shown in Fig.~\ref{RTfamily}. 

\subsection{Example: Pure AdS$_3$}

This subsection mirrors the discussion of the boundary modular Berry connection in the vacuum of a two-dimensional CFT. In Appendix~\ref{appendix:Vacuum} we identify the operator that generates modular parallel transport from boundary interval $\lambda$ to interval $\lambda + \delta\lambda$. In doing so, we only exploit the global conformal algebra $SO(2,1) \times SO(2,1)$. 

But this $SO(2,1) \times SO(2,1)$ is also the algebra of the Killing vector fields of AdS$_3$. In particular, equations~(\ref{adjoint1}-\ref{adjoint2}), (\ref{explicitv}) and (\ref{thevequation}) hold for the corresponding Killing vector fields. As a consequence, the Killing vector field that represents (\ref{explicitv}) is a solution of equation~(\ref{bulkmodHtransport}). We also know it has no zero mode component to be projected out
\begin{equation}
P_0^{\lambda}[\delta_{\lambda}\zeta_{\text{mod}}^M(x;\lambda)] = 0
\end{equation}
because---as was the case for operator $V_{\delta\lambda}$ in the boundary discussion---it too lives in eigenspaces of the adjoint action of $\zeta_{\rm mod}$ that are orthogonal to the 0-eigenspace. In summary, the Killing vector field that corresponds to $V_{\delta \lambda}$ generates the bulk modular parallel transport in pure AdS$_3$. 

To understand bulk modular parallel transport geometrically, consider an initial HRRT surface that is a diagonal of a static slice of AdS$_3$:
\begin{equation}
\lambda = (a^+, b^+, a^-, b^-) \equiv (\theta_L + t_L, \theta_R + t_R, \theta_L - t_L, \theta_R - t_R) = (-\pi/2, \pi/2, -\pi/2, \pi/2)
\label{speciallambda}
\end{equation}
The analysis for other initial geodesics is identical up to an overall AdS$_3$ isometry. The task is to interpret 
\begin{equation*}
\textrm{equation~(\ref{explicitv})} = \,\frac{1}{2\pi i} \left( -
 \partial_{a^+} K_+ +
\partial_{b^+} K_+ + 
\partial_{a^-} K_- -
\partial_{b^-} K_- \right),
\end{equation*}
the $SO(2,1) \times SO(2,1)$ algebra element that generates modular parallel transport, as an AdS$_3$ Killing vector field. In the representation~(\ref{lgenerators}), the action of (\ref{explicitv}) on the boundary is:
\begin{equation}
\! 
 \frac{1}{2}\! \frac{da^+}{d\lambda}\! \left( \sin x^+ + 1 \right)\!\partial_+ 
-\frac{1}{2}\! \frac{db^+}{d\lambda}\! \left( \sin x^+ - 1 \right)\!\partial_+ 
+\frac{1}{2}\! \frac{da^-}{d\lambda}\! \left( \sin x^- + 1 \right)\!\partial_- 
-\frac{1}{2}\! \frac{db^-}{d\lambda}\! \left( \sin x^- - 1 \right)\!\partial_-
\end{equation}  
Going to $\theta$ and $t$-coordinates on the boundary, this becomes:
\begin{align}
& \phantom{+} \frac{1}{2} \left( 
-\frac{d(b^+ - a^+)}{d\lambda} \sin (\theta + t) - \frac{d(b^- - a^-)}{d\lambda} \sin (\theta-t) 
+ \frac{d(a^++b^+ + a^- + b^-)}{d\lambda}
\right) \partial_\theta  \nonumber \\
& + \frac{1}{2} \left( 
-\frac{d(b^+ - a^+)}{d\lambda} \sin (\theta+t) + \frac{d(b^- - a^-)}{d\lambda} \sin (\theta-t)
+ \frac{d(a^++b^+ - a^- - b^-)}{d\lambda}
\right) \partial_t
\end{align}
Let us survey what this solution means in the bulk of AdS$_3$.

One option is to move from $\lambda$ to:
\begin{equation}
\lambda + \delta\lambda = (-\pi/2 + d\lambda, \pi/2 + d\lambda, -\pi/2 + d\lambda, \pi/2 + d\lambda)
\end{equation}
In this case, parallel transport is carried out by this global conformal symmetry:
\begin{equation}
V_{\delta \lambda}  = 2 \partial_\theta
\end{equation}
It is easy to see that the corresponding symmetry of AdS$_3$ is a global rotation about its center, which maps the geodesic $\lambda$ to $\lambda + \delta\lambda$. Mapping the special interval~(\ref{speciallambda}) to a general initial $\lambda$, we recognize the following rule of parallel transport:
\paragraph{Case 1:} If two geodesics live on a common $\mathbb{H}_2$ subspace of pure AdS$_3$ and intersect, bulk modular parallel transport is a rigid rotation about their intersection point which preserves their common $\mathbb{H}_2$. This rule for bulk modular parallel transport, dubbed `rotation without slipping,' was first explained in \cite{Czech:2017zfq}.

\bigskip 

Another case is to move from $\lambda$ to:
\begin{equation}
\lambda + \delta\lambda = (-\pi/2 + d\lambda, \pi/2 + d\lambda, -\pi/2 - d\lambda, \pi/2 - d\lambda)
\end{equation}
In this case, parallel transport is carried out by this global conformal symmetry:
\begin{equation}
V_{\delta \lambda}  = 2 \partial_t
\end{equation}
This is a rigid time translation in AdS$_3$. Once again, mapping the special interval~(\ref{speciallambda}) to a general initial $\lambda$, we recognize the following rule of parallel transport:
\paragraph{Case 2:} If two geodesics $\lambda$ and $\lambda + \delta\lambda$ live on a common AdS$_2$ subspace of pure AdS$_3$ and do not intersect, bulk modular parallel transport is a global time translation that preserves their common AdS$_2$. This time translation also preserves that timelike geodesic in AdS, which connects the points of closest approach between $\lambda$ and $\lambda + \delta\lambda$. 

\bigskip 

There are two other basic cases, which depend on the relative signs of $da^+/d\lambda$, $db^+/d\lambda$, $da^-/d\lambda$ and $db^-/d\lambda$. Altogether, these four basic cases span the four dimensions of kinematic space \cite{Czech:2016xec, Czech:2017zfq}. The most general case is of course a linear combination of the four.  Its detailed geometric meaning will be discussed in \cite{zizhi}.

Along any trajectory in the space of geodesics, parallel transport is generated by an AdS$_3$ isometry, which at each step maps geodesic $\lambda$ to geodesic $\lambda + \delta\lambda$. When we close a loop, we generate a finite AdS$_3$ isometry that maps the initial geodesic back to itself. Such isometries are spanned by the orthogonal boost and rigid translation along the said geodesic. Of course, we have reached the same conclusion in eqs.~(\ref{vacuumcurvature}): in the language of Sec.~\ref{cftvac}, the orthogonal boost is generated by $K_+ + K_-$ and the longitudinal translation by $K_+ - K_-$.

\section{The Proposal and Implications} 
\label{sec:conjecture}
In this paper, we proposed a link between the curvature of spacetime and the relations between modular Hamiltonians of the dual CFT state. Our key observation on the boundary is that the set of subregion modular Hamiltonians is endowed with a \emph{gauge symmetry}, consisting of rotating the basis of each $H_{\text{mod}}$ by a zero-mode transformation. The relative zero-mode frame is then promoted to a gauge connection with a non-vanishing curvature. This is a notion of curvature, which---as first recognized in \cite{Czech:2018kvg}---is directly associated to the entanglement pattern of the state. It can be studied by applying the ideas of Berry, Wilczek and Zee \cite{Berry:1984jv,Wilczek:1984dh} to the set of modular Hamiltonians.

\paragraph{Modular Berry holonomies as an entanglement measure} The characterization of multi-partite entanglement is a famously unsolved problem. Unlike two-partite entanglement, which is entirely characterized by the spectrum of the modular Hamiltonians, it is not known what quantities are sufficient to classify different forms of multi-partite entanglement.\footnote{Although there exist classification schemes that are customized to specific systems like qubits \cite{tangle, ghzw, 4qubit}.} Modular Berry holonomies are a promising quantity in this regard. One way in which one might probe multi-system entanglement is to group the systems into two sets and study how the resulting bipartite entanglement varies as the grouping evolves. This is a description of the modular Berry-Wilson loop. Because the focus of this paper is on holographic applications of the modular Berry connection, we leave an exploration of its uses for classifying entanglement to the future.

\paragraph{Modular Berry holonomies in holography}

In the bulk, modular flow admits a simple geometric description sufficiently close to the corresponding HRRT surface. This allowed us to translate the CFT rules of modular parallel transport to entanglement wedges and derive a bulk avatar of the modular Berry connection. Our main result is that for HRRT surfaces that satisfy condition (\ref{regimeofvalidity}) the modular Berry connection reduces to a geometric connection encoding the spin connection for the normal surface frame and the relative embedding of the internal coordinates. Its curvature is, therefore, a holographic probe of the bulk Riemann curvature. A somewhat different CFT Berry connection recently appeared in the discussion of holographic complexity \cite{Belin:2018fxe, Belin:2018bpg}, while the algebra of modular Hamiltonians was used for bulk reconstruction in \cite{Faulkner:2018faa, Kabat:2018smf}. Our feeling is that there is an overarching framework connecting these results to the ideas we presented here. 

\paragraph{Error correction and bulk locality.} Our proposed holographic relation between modular Berry curvature and bulk spacetime curvature hinges on the validity of the JLMS relation (\ref{JLMS}). The latter is the only bridge between our CFT and bulk discussions. The error correction framework for the AdS/CFT dictionary \cite{Harlow:2016vwg} clarified that the equivalence of bulk and boundary modular Hamiltonians (\ref{JLMS}) holds within the code subspace, namely the subspace of the CFT Hilbert space describing bulk low energy excitations about a specific background. It is therefore implicit in our construction that in the holographic context the $H_{\text{mod}}$ appearing in equations (\ref{CFTzeromode}) 
and (\ref{relativebasis}) 
is actually the restriction of the exact CFT modular Hamiltonian to the code subspace $H_{\text{mod}}=P_{\text{code}}H^{\text{exact}}_{\text{mod}}P_{\text{code}}$. 

The code subspace projection is more than just a technicality; it is directly responsible for endowing the boundary modular Hamiltonian with the right zero-mode algebra. In a typical CFT state, the symmetries of the modular Hamiltonian are either generated by $H_{\text{mod}}$ itself and the conserved global charges of the CFT, if any, or they are simple phase rotations of individual modular eigenstates. On the other hand, the existence of a \emph{local, semiclassical} bulk requires a set of zero-modes that generate the asymptotic symmetry group of the HRRT surface (\ref{bulkzeromodesexplicit}). The essential task of the projector $P_{\text{code}}$ is to introduce the correct group of approximate zero-modes. In the absence of any currently known, bulk-independent way for identifying the appropriate code subspaces in the boundary theory, the modular zero-mode algebra and the corresponding modular Berry holonomies can serve as a useful guiding principle.

\paragraph{On the role of soft modes.} There is an aspect of our story that played a supporting role in our main presentation but we believe deserves more attention. This is the new, to our knowledge, use of gravitational edge modes of subregions to probe the curvature of their embedding spacetime. Edge modes have been subject to a lot of recent studies due to their relation to soft theorems and the memory effect \cite{Strominger:2017zoo}, the construction of the physical phase space of subsystems in gauge theories  \cite{Donnelly:2016auv,Camps:2018wjf,Dong:2018seb,Kirklin:2019xug}, the definition of entanglement entropy  \cite{Donnelly:2014gva,Donnelly:2014fua}, and, more speculatively, to the black hole information problem \cite{Hawking:2016sgy, Haco:2018ske}. In our work, the relative edge mode frame of infinitesimally separated regions acquired a new physical interpretation as a gravitational connection with curvature that depends on the background spacetime. 

One moral of our treatment is that soft modes are unphysical, gauge degrees of freedom from the perspective of a given subregion but their holonomies contain physical geometric information. This informs the recent discussion regarding the physical significance of soft modes \cite{Bousso:2017dny, Bousso:2017rsx, Camps:2018wjf}. It will be illuminating to formulate our ideas more rigorously in the canonical formalism along the lines of \cite{Donnelly:2016auv}, where we believe they may offer a useful framework for describing surface translations. It is also worthwhile to apply them in backgrounds that are not asymptotically AdS.

We also learned that we can `implant' soft hair on the boundary of a subregion by transporting it around a closed loop. It is interesting to compare the latter with the more operational way of exciting soft modes by sending shockwaves that cross the boundary of the subregion \cite{Strominger:2017zoo}. Intuitively, the shockwaves of \cite{Strominger:2017zoo} can be thought of as the `experimental' protocol for shifting the location of the horizon---an idealized version of which is the transport problem we formulate in this paper. To construct a closed loop of surfaces we could apply two shockwaves along different directions, in two different orderings. The edge-mode holonomy in this setup measures the soft graviton component of the commutator of the two shockwaves. It would be interesting to understand this heuristic picture in detail. The appearance of shockwave commutators also suggests an intriguing possible relation to the physics of chaos \cite{Shenker:2013pqa,Maldacena:2015waa}.\footnote{We thank Beni Yoshida for this comment.}

\paragraph{Bulk gauge field holonomies.} An interesting playground for our ideas is the case of holographic CFTs with global symmetries. The conserved charges give rise to a new set of modular zero-modes, which are holographically mapped to the edge modes of the dual bulk gauge field. The relevant component of the modular Berry curvature should then be reflected in the local field strength of the gauge field along an HRRT surface.  This setup is, in a sense, simpler than the gravitational case we discussed in this work and could allow for more computations. For example, it would be an interesting exercise to repeat the computations we did for pure AdS$_3$ with a bulk gauge field turned on.

\paragraph{Gravitation and gauge field dynamics?} A particularly exciting question we leave for future study is whether our proposed perspective on the bulk gravitational and gauge connections can shed light on the emergence of their dynamics \cite{Harlow:2015lma}. An excitation of the CFT state gets imprinted on the modular Hamiltonians in its future causal cone and thus affects the modular Berry connection. As a result, the latter is ultimately promoted to a dynamical object. Whether the laws governing this evolution take a useful form, however, remains to be seen. It is worth noting that an appealing feature of our approach is that it treats all gauge fields, including gravity, on equal footing. All bulk holonomies have the same microscopic origin in the CFT: the entanglement pattern of the state as encoded in the relative bases of modular Hamiltonians. A dynamical law of the sort we speculate here would constitute a unified holographic description of gravitational and gauge interactions.

\section*{Acknowledgements}
We thank Xiao-Liang Qi, Lenny Susskind and Herman Verlinde for fruitful interactions and related work. We are also grateful to Vijay Balasubramanian, Gavin Brennen, Tom Faulkner, Matt Headrick, Hong Liu, Zizhi Wang, Jieqiang Wu, Beni Yoshida and Ellis Ye Yuan for interesting discussions and feedback. BC and LL thank the University of Pennsylvania, JdB thanks the Institute for Advanced Study (Princeton), and BC thanks MIT for hospitality while this work was completed. We all thank the organizers of the workshop ``Entanglement in Quantum Systems'' held at GGI Florence; BC, JdB and LL thank the organizers of workshop ``Modern Techniques for CFT and AdS'' held at MITP Mainz; BC thanks the organizers of ``Workshop on Black Holes and Holography'' held at TSIMF Sanya and of conference ``Tensor Networks: From Simulations to Holography II'' held at AEI Postdam and DESY Zeuthen. BC is supported by the Startup Fund from Tsinghua University and by the Young Thousand Talents Program. LL is supported by the Pappalardo Fellowship.

\appendix
\section{Berry connection}
\label{appendix:Berry}
Consider a family of normalized pure states $\rho(\lambda)= |\psi(\lambda)\rangle \langle \psi(\lambda)|$. Each state is invariant under the transformation $U(\lambda)=\exp\Big(i \theta(\lambda) |\psi(\lambda)\rangle \langle \psi(\lambda)|\Big)$, which simply rotates the vector $|\psi(\lambda)\rangle$ by a phase $\theta(\lambda)$. The operators $U(\lambda)$ are therefore the modular zero modes in this simple example.

The variation of the state under an infinitesimal change of $\lambda$ is
\begin{equation}
\partial_{\lambda}\rho = \left( \partial_\lambda |\psi\rangle\right) \langle \psi| +|\psi\rangle \left(\partial_{\lambda} \langle \psi|\right) = [V,\rho] \label{purestatevariation}
\end{equation}
where we defined the anti-Hermitian operator 
\begin{equation}
V= \left( \partial_\lambda |\psi\rangle\right) \langle \psi| -|\psi\rangle \left(\partial_{\lambda} \langle \psi|\right) 
\end{equation}
This is clearly not unique since any addition of zero-modes to $V$ respects equation (\ref{purestatevariation}). This reflects our freedom to independently rotate the phases of $|\psi(\lambda)\rangle$ and $|\psi(\lambda+\delta \lambda)\rangle$.

According to (\ref{modularconnection}), the modular Berry connection is the projection of $V(\lambda)$ onto the zero modes of $\rho(\lambda)$ which, using the projector (\ref{zeromodeprojectionHS}), reads:
\begin{equation}
\Gamma= P_0^{\lambda}[V(\lambda)]= \Big(\langle \psi|\partial_{\lambda}\psi\rangle - \langle \partial_{\lambda}\psi|\psi\rangle \Big) |\psi\rangle\langle\psi |
\end{equation}
This is the familiar Berry connection \cite{Berry:1984jv,Wilczek:1984dh}.

\section{Modular connection for CFT vacuum}
\label{appendix:Vacuum}
The two-sided modular Hamiltonian for an interval in the CFT vacuum can be written in terms of the conformal generators as $H_{\text{mod}}=K_+ + K_-$, with:
\begin{align}
K_+ & = s_1 L_1 + s_0L_0 +s_{-1} L_{-1} \\
K_- & = t_1 \bar{L}_1 + t_0 \bar{L}_0 +t_{-1} \bar{L}_{-1}.
\end{align}
The coefficients $s_i, t_i$ are determined, up to an overall multiplicative constant, by the requirement that the generators $K_+$ and $K_-$ preserve the left-moving and right-moving null coordinates of the interval endpoints $(a^+,b^+)$ and $(a^-,b^-)$, respectively. Working in the representation
\begin{equation}
L_{-1} = i e^{-ix^+} \partial_+ \qquad {\rm and} \qquad 
L_0 = i \partial_+ \qquad {\rm and} \qquad
L_1 = i e^{ix^+} \partial_+,
\label{lgenerators}
\end{equation}
with an identical action of the $\bar{L}_i$s on the $x^-$ null coordinate, we find:
\begin{equation}
\begin{tabular}{rlp{1cm}rl}
$s_1$ & $= \frac{2 \pi \cot (b^+-a^+)/2}{e^{i a^+}+e^{ib^+}}$ & & $t_1$ & $= -\frac{2 \pi \cot (b^--a^-)/2}{e^{i a^-}+e^{ib^-}}$ \\
$s_0$ & $=-2\pi \cot (b^+-a^+)/2$ & & $t_0$ & $= 2\pi \cot (b^--a^-)/2 $\\
$s_{-1}$ & $= \frac{2\pi \cot (b^+-a^+)/2}{e^{-i a^+}+e^{-ib^+}}$ & & $t_{-1}$ & $=-\frac{2\pi \cot (b^--a^-)/2}{e^{-i a^-}+e^{-ib^-}}$
\end{tabular}
\label{parameters}
\end{equation}
We found the overall magnitude of $H_{\rm mod}$ by demanding that $\exp( -H_{\rm mod}/2)$---a finite $SO(2,1) \times SO(2,1)$ transformation---map an interval to its complement.

The generator of modular parallel transport is defined by the conditions:
\begin{align}
\partial_{a^+} K_+ &= [V_{\delta a^+}, K_+] \label{AconformalV1} \\
P_0[V_{\delta a^+}]&=0 \label{AconformalV2}
\end{align}
In the vacuum of a two-dimensional CFT, any single-interval modular Hamiltonian can be mapped to any other using conformal transformations. This is the reason for the absence of the spectrum changing operator appearing on the left hand side of the general equation (\ref{Vtransportdefinition}). The same fact guarantees that $V_{\delta a^+}$ is an element of the conformal algebra, so it is a linear combination of the generators (\ref{lgenerators}). 

To find $V_{\delta a^+}$ explicitly, it is convenient to decompose the conformal algebra into eigenoperators of the adjoint action of the modular Hamiltonian:
\begin{equation}
[K_+, E_{\kappa} ] = \kappa E_{\kappa} \label{eigenoperators}
\end{equation}
The three solutions of equation~(\ref{eigenoperators}) are:
\begin{align}
[K_+, K_+] & = 0 \nonumber \\
[K_+, \partial_{a^+} K_+] & = -2 \pi i \,\partial_{a^+} K_+ 
\label{adjoint1} \\
[K_+, \partial_{b^+} K_+] & = +2 \pi i \,\partial_{b^+} K_+
\label{adjoint2}
\end{align}
This immediately implies eq.~(\ref{horizontalvacuum}), i.e.:
\begin{equation}
V_{\delta a^+}=\frac{1}{2\pi i}\, \partial_{a^+} K_+
\end{equation}
Eq.~(\ref{AconformalV2}) is automatically satisfied because $\partial_{a^+} K_+$ and $K_+$ live in orthogonal eigenspaces of the eigenvalue equation~(\ref{eigenoperators}).
 
 
It is easy to consider a more general direction in kinematic space (space of CFT intervals.) Say we go from $H_{\rm mod}(\lambda)$ (here $\lambda = (a^+, b^+, a^-, b^-)$) to $\lambda + \delta\lambda$. The change in the modular Hamiltonian is: 
\begin{equation}
\partial_\lambda H_{\rm mod} \equiv 
\big(\partial a^+ / \partial\lambda)\, \partial_{a^+} K_+ + 
\big(\partial b^+ / \partial\lambda)\, \partial_{b^+} K_+ + 
\big(\partial a^- / \partial\lambda)\, \partial_{a^-} K_- +
\big(\partial b^- / \partial\lambda)\, \partial_{b^-} K_- \,.
\end{equation}
The operator
\begin{equation}
V_{\delta \lambda} = \frac{1}{2\pi i} \left( 
 \frac{\partial a^+}{\partial\lambda} \partial_{a^+} K_+ 
- \frac{\partial b^+}{\partial\lambda} \partial_{b^+} K_+ 
- \frac{\partial a^-}{\partial\lambda} \partial_{a^-} K_- 
+ \frac{\partial b^-}{\partial\lambda} \partial_{b^-} K_- 
\right)
\label{explicitv}
\end{equation}
solves 
\begin{equation}
[V_{\delta \lambda}, H_{\rm mod}] = \partial_\lambda H_{\rm mod}.
\label{thevequation}
\end{equation}
It also satisfies (\ref{AconformalV2}) because it lives outside the zero-eigenspace of $[H_{\rm mod}, E_\kappa] = \kappa E_\kappa$, the latter being generated by $K_+$ and $K_-$. Therefore, (\ref{explicitv}) is the generator of modular parallel transport.

\section{Solution to equation (\ref{bulkmodHtransport})}
\label{appendix:solution}

Consider an entanglement wedge $\lambda$ and a coordinate system $x^M=(x^{\alpha}, y^i)$ in the neighborhood of its RT surface. $x^{\alpha}$ denotes distances along two directions orthogonal to the RT surface and $y^i$ is a choice of internal surface coordinates. 

Since we are ultimately interested in comparing the frames of two nearby extremal surfaces, the form of the metric in the vicinity of the RT surface is important. It is convenient to introduce normal geodesic coordinates $\sigma^M= (\sigma^a (x), y^i)$, where $\sigma^a(x)\eta_{ab} \sigma^b(x)$ measures the geodesic distance of a nearby point $x$ from the minimal surface and $\frac{\sigma^a}{\sigma}$ is the unit tangent vector to the same geodesic at its starting point on the surface. In an expansion around the surface, this coordinate system is: 
\begin{align}
x^M(\sigma^{\alpha},y^i)&= \sigma^M -\frac{1}{2}\Gamma^M_{\alpha\beta}(y) \sigma^{\alpha}\sigma^{\beta} +\mathcal{O}(\sigma^3) \label{geodesicframe}
\end{align}
The advantage of the $\sigma$-coordinates is that they set the components $\Gamma^M_{\alpha\beta}$ of the Christofel connection to zero, so they constitute the analog of the local inertial frame for a surface. 

For as long as we focus on a small neighborhood of the RT surface $( \sigma^{+}K_{ij|+}, \,\sigma^{-}K_{ij|-}\ll 1)$ the action of the modular Hamiltonian is expected to be local and, therefore, it can be described by a vector field $\zeta_{(\lambda)}^M(\sigma)$ generating a geometric flow. About the surface, the modular flow generator has the form:
\begin{align}
\zeta_{(\lambda)}^a &= 2\pi \epsilon^{a}\,_b \sigma^b \nonumber\\
\zeta_{(\lambda)}^i &= 0 \label{modboost}
\end{align}


\paragraph{The $\lambda$-derivative of the modular boost.} Consider now a nearby entanglement wedge $\lambda+\delta\lambda$ whose RT surface is separated from that of $\lambda$ by $\delta \sigma^{\alpha}(y_i)$ in the orthogonal directions. Its modular boost generator $\tilde{\zeta}_{(\lambda+\delta\lambda)}^M$ will have the same form (\ref{modboost}) in the normal frame of the new wedge. Let $n_{a}\,^M(\tilde{y} ;\lambda+\delta\lambda)$ ($a=0,1$) be two orthonormal vectors at every point $\tilde{y}^i$ on the HRRT surface of $\lambda+\delta\lambda$ and denote by $s^a$ distances along $n_a\,^M$. Then the map $\left(\sigma^a(s^a,\tilde{y}^i),y^i(s^a,\tilde{y}^i)\right)$, at first non-trivial order in the separation of the two surfaces, is:
\begin{align}
\sigma^a&= \delta \sigma^a(\tilde{y}) + s^a +\delta n_b\,^a s^b +\mathcal{O}(s^2, \delta \sigma^2, \delta n^2)\nonumber\\
y^i&= \tilde{y}^i +\delta n_b\,^i s^b +\mathcal{O}(s^2, \delta \sigma^2, \delta n^2) \label{sigmatos}
\end{align}
Here $\delta n_a\,^M \equiv n_a\,^M(\lambda+\delta\lambda) - n_a\,^M(\lambda)$ and we have used the fact that, in the orthonormal gauge we are using, the normal vectors on the $\lambda$ surface are $n_a\,^M (\lambda)= \delta^M_a$.

It is important to note that the choice of normal coordinates $s^a$ is not unique, since any local Lorentz boost on the orthogonal plane 
\begin{equation}
\delta n_b\,^a \rightarrow \delta n_b\,^a +\omega(\tilde{y}) \epsilon_b\,^a \label{nambiguity}
\end{equation}
will yield an equally acceptable pair of normal directions. There is, therefore, an ambiguity in the map between the normal frames of two nearby minimal surfaces. This ambiguity will be important in what follows.

Since the vector field $\tilde{\zeta}_{(\lambda+\delta\lambda)}^M$ has the form (\ref{modboost}) in the $s^a$-coordinates, we can use transformation (\ref{sigmatos}) to map it back to the $\sigma$-coordinates and compute the difference of the two modular boost generators:
\begin{align}
\delta_{\lambda} \zeta^M&= -\delta^M_a \epsilon^a\,_b \delta\sigma^b - \delta^M_a \epsilon^a\,_b \delta n_c\,^b \sigma^c + \delta n_a\,^M \epsilon^a\,_b \sigma^b   \nonumber\\
&+\mathcal{O}(s^2, \delta \sigma^2, \delta n^2) \label{deltalambdazeta}
\end{align}

\paragraph{Zero-mode component of (\ref{deltalambdazeta}).}The next step is to compute the zero mode component of $\delta_{\lambda} \zeta^M$ and subtract it to obtain an equation for the Lie bracket of $\xi$ with $\zeta_{\lambda}$. Applying the projector (\ref{bulkconnection}) to the right hand side of (\ref{deltalambdazeta}), we find:
\begin{align}
\Omega \left(\delta_{\lambda}\zeta \right) &= -\frac{1}{2} \epsilon^{ab}\partial_a \left(\eta_{bc}\delta_{\lambda}\zeta^b \right) \Big|_{\sigma^a=0}= 0 \\
Z^i \left(\delta_{\lambda}\zeta \right) &= \delta_{\lambda}\zeta^i \Big|_{\sigma^a=0}=0
\end{align}
Equation (\ref{deltalambdazeta}) contains no zero mode components, so no extra subtraction is necessary.



\paragraph{The bulk modular connection.} By plugging the result (\ref{deltalambdazeta}) into equation (\ref{bulkmodHtransport}) we obtain an equation for the diffeomorphism $\xi$ that can be straightforwardly solved to get
\begin{equation}
\xi^M= -\delta^M_a\,\delta\sigma^a -\delta n_a\,^M \,\sigma^a \label{xisolution}
\end{equation}
The solution (\ref{xisolution}) is not unique, because the vector Lie bracket $[\xi, \zeta]$ has a kernel. The family of solutions of (\ref{bulkmodHtransport}) are related to (\ref{xisolution}) (and each other) by:
\begin{equation}
\xi^M \rightarrow  \xi^M +\omega(\tilde{y}^i) \delta^M_a\, \epsilon^a\,_b \,\sigma^b  +\delta^M_i \, \zeta_0^i(\tilde{y}_i) \label{xiambiguity}
\end{equation}
As discussed in the main text, this is simply an addition of modular zero modes. The first term, corresponding to a spatially varying boost along the orthogonal RT surface directions, can be absorbed in the ambiguity (\ref{nambiguity}) in the local choice of normal vectors on the RT surface of $\lambda+\delta\lambda$. The second term, in turn, allows the internal coordinate systems on $\lambda$ and $\lambda+\delta\lambda$ to be related by an infinitesimal element of the $d-2$-dimensional diffeomorphim subgroup.

It instructive to transform the result (\ref{xisolution}) back to the general normal gauge $x^M$ using (\ref{geodesicframe}). The computation yields the following general solution (up to zero modes):
\begin{equation}
\xi^M = -\delta^M_a\,\delta\sigma^a -\left(\delta n_a\,^M +\Gamma^M_{ab}\,\delta\sigma^b \right)\,x^a   \label{xisolutiongeneral}
\end{equation}


\end{document}